\documentstyle[12pt,psfig]{article}

\renewcommand{\theequation}
{\arabic{section}.\arabic{equation}}

\makeatletter
\def\eqnarray{ \stepcounter{equation} \let\@currentlabel=\theequation
 \global\@eqnswtrue
 \global\@eqcnt\z@
 \tabskip\@centering
 \let\\=\@eqncr
 $$\halign to \displaywidth\bgroup\@eqnsel\hskip\@centering
 $\displaystyle\tabskip\z@{##}$&\global\@eqcnt\@ne
 \hfil$\displaystyle{{}##{}}$\hfil
 &\global\@eqcnt\tw@$\displaystyle\tabskip\z@{##}$\hfil
 \tabskip\@centering&\llap{##}\tabskip\z@\cr}
\makeatother

\makeatletter
\def\@arrayacol{\edef\@preamble{\@preamble \hskip .5\arraycolsep}}
\def\array{\let\@acol\@arrayacol \let\@classz\@arrayclassz
\let\@classiv\@arrayclassiv \let\\\@arraycr\def\@halignto{}\@tabarray}
\makeatother

\makeatletter
\newcounter{subeqncnt}
\def\thesubeqncnt{\alph{subeqncnt}}
\def\subequations{\begingroup%
   \stepcounter{equation}\edef\@tempa{\theequation}%
   \let\c@equation\c@subeqncnt\c@subeqncnt\z@
   \edef\theequation{\@tempa\noexpand\thesubeqncnt}}

\makeatother

\newcommand{\be}{\begin{equation}}
\newcommand{\ee}{\end{equation}}

\newcommand{\beqa}{\begin{eqnarray}}
\newcommand{\eeqa}{\end{eqnarray}}
\newcommand{\nn}{\nonumber}

\newcommand{\eqref}[1]{(\ref{#1})}

\def\CF {{\cal F}}

\def\CL {{\cal L}}

\def\CO {{\cal O}}

\def\CV {{\cal V}}

\begin{document}

\setlength{\baselineskip}{7mm}
\begin{titlepage}
\begin{flushright}
{\tt NRCPS-HE-56-08} \\
August, 2008
\end{flushright}

\vspace{1cm}

\begin{center}
{\Large \it  Connection Between   \\
\vspace{1cm}
Non-Abelian Tensor Gauge Fields
and Open Strings
}

\vspace{1cm}

{ {George  Savvidy  }\footnote{savvidy(AT)inp.demokritos.gr}}

\vspace{1cm}

 {\it Institute of Nuclear Physics,} \\
{\it Demokritos National Research Center }\\
{\it Agia Paraskevi, GR-15310 Athens, Greece}

\end{center}

\vspace{1cm}

\begin{abstract}
We compare the structure of the tree level
scattering amplitudes in non-Abelian tensor gauge field theory and in open string
theory with Chan-Paton charges. We limit ourselves considering only lower rank
tensor fields in both theories. We identify the symmetric  and antisymmetric components of the
second rank tensor gauge field  with the string excitations
on the third and forth levels. In the process of this identification we have been
selecting only those parts of the tree level scattering amplitudes in the open string theory
which have dimensionless coupling constants in four dimensions. It seems that
this subclass of tree level scattering amplitudes may provide important information
about the structure of the open string theory and most probably
it is equivalent to the non-Abelian tensor gauge field theory.

\end{abstract}

\end{titlepage}

\section{\it Introduction }

An infinite tower of particles of high spin naturally
appears in the spectrum of different string theories. In the low
energy limit the massless states of the open string theory with Chan-Paton charges
can be identified with the Yang-Mills gauge quanta \cite{Neveu:1971mu,Green:1987sp,Polchinski:1998rq}.
It is also expected that in the tensionless limit
or, what is equivalent, at high energy and fixed angle scattering
the string spectrum becomes effectively massless \cite{Gross:1988ue,
Gross:1987kz,Gross:1987ar,Witten:1988zd,Amati:1988tn,Lindstrom:1990qb,
Mende:1992pm,Mende:1989wt,DeVega:1992tm, Mende:1994wf, Moore:1993ns,
Savvidy:2003fx,Giddings:2007bw,Bakas:2004jq,Savvidy:2005fe}. It is of great importance
to identify these states with the spectrum of some Lagrangian quantum field theory
\cite{Witten:1985cc,Thorn:1985fa,Siegel:1985tw,Siegel:1988yz,Arefeva:1989cp,
Taylor:2003gn,Taylor:2006ye}.

One can imagine that these massless states are combined into the infinite tower
of tensor gauge fields and one could guess that the possible solution could be found
by the extension of the Yang-Mills gauge symmetry \cite{yang} to non-Abelian
tensor gauge fields.  This possibility was suggested recently in
\cite{Savvidy:2005fi,Savvidy:2005zm,Savvidy:2005ki}.
Recall that non-Abelian gauge fields are defined as
rank-(s+1) tensor gauge fields
$
A^{a}_{\mu\lambda_1 ... \lambda_{s}}
$
\footnote{Tensor gauge fields
$A^{a}_{\mu\lambda_1 ... \lambda_{s}}(x),~~s=0,1,2,...$
are totally symmetric with respect to the
indices $  \lambda_1 ... \lambda_{s}  $. {\it A priori} the tensor fields
have no symmetries with respect to the first index  $\mu$.
In particular, we have
$A^{a}_{\mu\lambda}\neq A^{a}_{\lambda\mu}$ and
$A^{a}_{\mu\lambda\rho}=A^{a}_{\mu\rho\lambda} \neq A^{a}_{\lambda\mu\rho}$.
The adjoint group index $a=1,...,N^2 -1$
in the case of $SU(N)$ gauge group. }
and that one can construct infinite series of forms
$ {{\cal L}}_{s}~( s=1,2,..)$  which are
invariant with respect to the  extended gauge transformations
\cite{Savvidy:2005fi,Savvidy:2005zm,Savvidy:2005ki}.
These forms $ {{\cal L}}_{s}$ are quadratic in the field strength tensors
$G^{a}_{\mu\nu,\lambda_1 ... \lambda_s}$ and the general Lagrangian is
an infinite sum of these forms (\ref{Lagrangian}).

The resulting gauge invariant Lagrangian
defines {\it cubic and quartic interactions} with
{\it dimensionless coupling constant}  between charged gauge quanta
\cite{Savvidy:2005fi,Savvidy:2005zm,Savvidy:2005ki}
$$
A^{a}_{\mu},~~~~~ A^{a}_{\mu\lambda_1},~~~~~A^{a}_{\mu\lambda_1\lambda_2},~~~~~.....
$$
carrying a spin larger than one. Note that all these non-Abelian tensor gauge bosons have
the same isotopic charges as the vector gauge boson $A^{a}_{\mu}$. The gauge invariant
Lagrangian describing dynamical tensor gauge bosons of all ranks
has the form
\be\label{Lagrangian}
{\cal L} ~= ~
{{\cal L}}_{YM} + {\cal L}_2 +g_{3 } {\cal L}_3  +....,
\ee
where ${{\cal L}}_{YM}$ is the Yang-Mills Lagrangian.
For the lower-rank tensor gauge fields the Lagrangian has the following form
\cite{Savvidy:2005fi,Savvidy:2005zm,Savvidy:2005ki}:
\beqa\label{totalactiontwo}
{{\cal L}}_{YM} =& -&{1\over 4}G^{a}_{\mu\nu}
G^{a}_{\mu\nu},\nn\\
{{\cal L}}_2  =&-&
{1\over 4}G^{a}_{\mu\nu,\lambda}G^{a}_{\mu\nu,\lambda}
-{1\over 4}G^{a}_{\mu\nu}G^{a}_{\mu\nu,\lambda\lambda}  \\
&+&{1\over 4}G^{a}_{\mu\nu,\lambda}G^{a}_{\mu\lambda,\nu}
+{1\over 4}G^{a}_{\mu\nu,\nu}G^{a}_{\mu\lambda,\lambda}
+{1\over 2}G^{a}_{\mu\nu}G^{a}_{\mu\lambda,\nu\lambda},\nn
\eeqa
where the generalized field strength tensors are:
\beqa\label{tensors}
G^{a}_{\mu\nu} &=&
\partial_{\mu} A^{a}_{\nu} - \partial_{\nu} A^{a}_{\mu} +
g f^{abc}~A^{b}_{\mu}~A^{c}_{\nu}, \nn\\
G^{a}_{\mu\nu,\lambda} &=&
\partial_{\mu} A^{a}_{\nu\lambda} - \partial_{\nu} A^{a}_{\mu\lambda} +
g f^{abc}(~A^{b}_{\mu}~A^{c}_{\nu\lambda} + A^{b}_{\mu\lambda}~A^{c}_{\nu} ~), \\
G^{a}_{\mu\nu,\lambda\rho} &=&
\partial_{\mu} A^{a}_{\nu\lambda\rho} - \partial_{\nu} A^{a}_{\mu\lambda\rho} +
g f^{abc}(~A^{b}_{\mu}~A^{c}_{\nu\lambda\rho} +
 A^{b}_{\mu\lambda}~A^{c}_{\nu\rho}+A^{b}_{\mu\rho}~A^{c}_{\nu\lambda}
 + A^{b}_{\mu\lambda\rho}~A^{c}_{\nu} ~) .\nn
\eeqa
The Lagrangian forms $ {{\cal L}}_{s}$
for higher-rank  fields can be found in the previous
publications \cite{Savvidy:2005fi,Savvidy:2005zm,Savvidy:2005ki}.
The above expressions define interacting gauge field theory with infinite
many non-Abelian tensor gauge fields. Not much is known about physical properties of such gauge
field theory and this article is one in the series of articles devoted to
its studies
\cite{Savvidy:2005at,Barrett:2007,Guttenberg:2008zn,Konitopoulos:2008vv,Konitopoulos:2008bd}.

In the present paper we shall focus our attention on the lower-rank
tensor gauge field $A^{a}_{\mu\lambda}$, which
decomposes in this theory to the symmetric tensor $T_S$ of helicity two and antisymmetric
tensor $T_A$ of helicity zero charged gauge bosons \cite{Savvidy:2005ki}.
The Feynman rules for these propagating modes and their interaction vertices
can be extracted from the above Lagrangian (\ref{totalactiontwo})
\cite{Savvidy:2005ki} and are reviewed in the next section. These Feynman rules
allow, in particular, to calculate tree-level scattering amplitudes for the production of
tensor gauge boson in annihilation processes
\cite{Konitopoulos:2008vv,Konitopoulos:2008bd}.

\begin{figure}
\centerline{\hbox{\psfig{figure=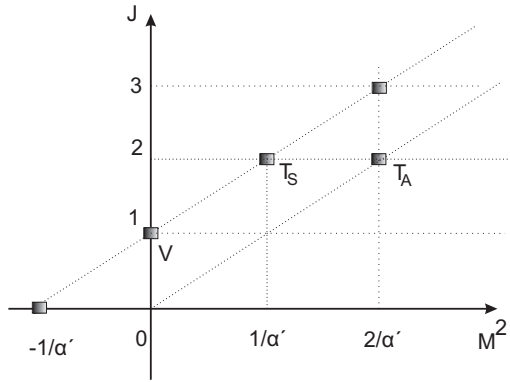,height=5cm,angle=0}}}
\caption[fig1]{The first excited levels of the open bosonic string. The state
at the third level is a symmetric, traceless, rank-2 tensor $T_S$ of $SO(D-1)$. At the
fourth excited level there is an antisymmetric rank-2 tensor $T_A$, of $SO(D-1)$. If the
open string carries Chan-Paton charges at its endpoints then these
excited states have the same isotopic charges as the massless vector boson V on the
first excited level.}
\label{fig1}
\end{figure}

It is an interesting question if this non-Abelian tensor gauge field theory have
any connection with the open string theory with Chan-Paton
charges \cite{Paton:1969je}\footnote{This question was raised to the author by Costas Bachas and
initiated this investigation. }.
In the spectrum of the open string theory with Chan-Paton charges there is a
massless vector gauge
boson $V$ on the second level and there are rank-two massive tensor bosons $T_S$
on the third level and $T_A$ on the forth level   carrying  the same isotopic charges as the
vector boson V. These states are depicted schematically on the Fig.\ref{fig1}
as $T_S$ and $T_A$.
The emission vertices for these states are defined as follows
\cite{Green:1987sp,Polchinski:1998rq,Kawai:1985xq}:
\beqa
&e_{\alpha}(k):\dot{X}^{\alpha} e^{ik  X}: ~~~~~~~~~~~~~  &\alpha^{'} k^2 = 0\nn\\
&\varepsilon_{\alpha\alpha^{'}}(k):\dot{X}^{\alpha}\dot{X}^{\alpha^{'}} e^{ik  X}: ~~~~~~~~~~~~~
&\alpha^{'} k^2 = -1\nn\\
&\zeta_{\alpha\alpha^{'}}(k){1\over 2}:(\ddot{X}^{ \alpha}\dot{X}^{\alpha^{'} }-
\ddot{X}^{\alpha^{'} }\dot{X}^{\alpha })  e^{ik  X} :~~~~~~~~~~&\alpha^{'} k^2 = -2.
\eeqa
They allow to calculate different tree level scattering amplitudes
for these tensor bosons.

{\it Our intension in this article is to compare the tree-level scattering amplitudes
of the second rank tensor gauge bosons in non-Abelian tensor gauge field theory and tree-level
scattering amplitudes of the tensor bosons in open string theory with Chan-Paton charges.}

Our aim is two-fold: first, to review the general
structure of the interaction vertices in non-Abelian tensor gauge field theory
\cite{Savvidy:2005fi,Savvidy:2005zm,Savvidy:2005ki} and, second, to calculate
similar tree-level scattering amplitudes in the open string
theory in order to compare their structures. By "similar" tree-level scattering amplitudes
we mean the special selection of amplitudes of the
open string theory when the corresponding amplitudes in four-dimensional space-time
have {\it dimensionless coupling constants}.  These amplitudes should have
only one space-time derivative in the case
of triple interaction vertices and have no derivatives in the case
of quartic interactions. We are choosing this subclass of interaction amplitudes
of open string theory because this subsector of amplitudes  with dimensionless
coupling constants naturally appears in the low-energy limit  when
one can ignore the higher derivative terms.
Only this subclass of vertices appears in non-Abelian
tensor gauge field theory \cite{Savvidy:2005fi,Savvidy:2005zm,Savvidy:2005ki}.

The open string tree-level scattering amplitudes are defined only on the mass-shell
 \cite{Green:1987sp,Polchinski:1998rq}.
Therefore in order to compare them with the tensor gauge field theory
vertices we have to project them to the mass-shell. We have found
that after projection the scattering amplitudes
with dimensionless coupling constant for the lower-rank tensor fields in
these theories coincide.
This result tells that most probably the subsector
of the open string theory with dimensionless coupling constants is equivalent
to the non-Abelian tensor gauge field theory.

\section{{\it Feynman Rules for Non-Abelian Tensor Gauge Fields}}

We shall describe here  the interaction vertices
of the vector and tensor gauge bosons in non-Abelian tensor gauge field theory
\cite{Savvidy:2005fi,Savvidy:2005zm,Savvidy:2005ki}. For that
let us recapitulate the
construction of the corresponding Feynman rules \cite{Savvidy:2005ki}.
Because the Lagrangian (\ref{Lagrangian})
is quadratic in field strength tensors $(G)^2 = (dA + g [A,A])^2$, it
allows only {\it cubic and quartic  interactions} with
{\it dimensionless coupling constants}:
$$
g~ [A,A]dA  ,~~~~~~~~~~g^2~ [A,A][A,A] .
$$
In particular,
the interaction of the Yang-Mills vector bosons with
the charged tensor gauge bosons described by the Lagrangian
(\ref{totalactiontwo}) are of this type. The  Lagrangian (\ref{totalactiontwo})
can be represented in the
form of polynomial in Yang-Mills vector field $A^{a}_{\alpha}$ and
tensor gauge field of the second  rank $A^{a}_{\alpha\acute{\alpha}}$
\cite{Savvidy:2005ki}:
\beqa
{{\cal L}}_2 =  {1 \over 2} A^{a}_{\alpha\acute{\alpha}}
H_{\alpha\acute{\alpha}\gamma\acute{\gamma}} A^{a}_{\gamma\acute{\gamma}}+
{1 \over 2!}\CV^{abc}_{ \alpha \acute{\alpha} \beta \gamma\acute{\gamma} }
A^{a}_{\alpha\acute{\alpha}} A^{b}_{\beta}A^{c}_{\gamma\acute{\gamma}}+
{1 \over 2! 2!}
\CV^{abcd}_{ \alpha \beta \gamma\acute{\gamma} \delta\acute{\delta }}
A^{a}_{\alpha } A^{b}_{\beta} A^{c}_{\gamma\acute{\gamma}}
A^{d}_{\delta\acute{\delta}}+...
\eeqa
It has the kinetic term $AHA$ for the tensor gauge field  and the interaction vertices between two tensors
and a vector, the VTT-vertex $\CV^{abc}_{ \alpha \acute{\alpha} \beta \gamma\acute{\gamma} }$ and
two tensors and two vectors, the VVTT-vertex
$\CV^{abcd}_{ \alpha \beta \gamma\acute{\gamma} \delta\acute{\delta }}$.
 The kinetic operator of the Lagrangian is
\beqa\label{basickineticterm}
H_{\alpha\acute{\alpha}\gamma\acute{\gamma}}(k)=
(-\eta_{\alpha\gamma}\eta_{\acute{\alpha}\acute{\gamma}}
+{1 \over 2}\eta_{\alpha\acute{\gamma}}\eta_{\acute{\alpha}\gamma}
+{1 \over 2}\eta_{\alpha\acute{\alpha}}\eta_{\gamma\acute{\gamma}})k^2
+\eta_{\alpha\gamma}k_{\acute\alpha}k_{\acute{\gamma}}
+\eta_{\acute\alpha \acute{\gamma}}k_{\alpha}k_{\gamma}\nn\\
-{1 \over 2}(\eta_{\alpha\acute{\gamma}}k_{\acute\alpha}k_{\gamma}
+\eta_{\acute\alpha\gamma}k_{\alpha}k_{\acute{\gamma}}
+\eta_{\alpha\acute\alpha}k_{\gamma}k_{\acute{\gamma}}
+\eta_{\gamma\acute{\gamma}}k_{\alpha}k_{\acute\alpha})
\eeqa
and it is symmetric under simultaneous interchange of the indices
$\alpha \leftrightarrow \acute{\alpha}$ and
$\gamma \leftrightarrow \acute{\gamma}$, but it is not symmetric
with respect to a single interchange
$\alpha \leftrightarrow \acute{\alpha}$ or
$\gamma \leftrightarrow \acute{\gamma}$,
because the tensor field
$A^{a}_{\alpha\acute{\alpha}}$
is not a symmetric tensor. It is also fully gauge invariant operator
$
k_{\alpha}H_{\alpha\acute{\alpha}\gamma\acute{\gamma}}=0,~
k_{\acute{\alpha}}H_{\alpha\acute{\alpha}\gamma\acute{\gamma}}=0,
$
therefore there is no negative norm states
in the spectrum \cite{Savvidy:2005fi,Savvidy:2005zm,Savvidy:2005ki}.
It describes the propagation of massless particles with helicities two
and zero. Indeed, when $k_{\mu}$ is aligned along the third axis,
$k_{\mu}= (k,0,0,k)$, the equation
\be
H_{\alpha\acute{\alpha}\gamma\acute{\gamma}}(k) f^{\gamma\acute{\gamma}}(k)=0
\ee
has three independent solutions of the helicity two
and zero
\beqa\label{wavefunctions}
\varepsilon^{1}_{\alpha\acute{\alpha}}={1\over \sqrt{2}}
\left( \begin{array}{llll}
  0,0,~~0,0\\
  0,1,~~0,0\\
  0,0,-1,0\\
  0,0,~~0,0
\end{array} \right), \varepsilon^{2}_{\alpha\acute{\alpha}}={1\over \sqrt{2}}
\left( \begin{array}{ll}
  0,0,0,0\\
  0,0,1,0\\
  0,1,0,0\\
  0,0,0,0
\end{array} \right),
\zeta_{\alpha\acute{\alpha}}={1\over \sqrt{2}}
\left( \begin{array}{ll}
  0,~~0,0,0\\
  0,~~0,1,0\\
  0,-1,0,0\\
  0,~~0,0,0
\end{array} \right),\nn\\
\eeqa
with the property that
$
\varepsilon^{1}_{\gamma\acute{\gamma}}\varepsilon^{1}_{\lambda\acute{\lambda}}  +
\varepsilon^{2}_{\gamma\acute{\gamma}}\varepsilon^{2}_{\lambda\acute{\lambda}} \simeq
{1\over 2}(\eta_{\gamma\lambda}\eta_{\acute{\gamma}\acute{\lambda}}
+\eta_{\gamma\acute{\lambda}}\eta_{\acute{\gamma}\lambda}
-\eta_{\gamma\acute{\gamma}}\eta_{\lambda\acute{\lambda}})
$
and
$
\zeta_{\gamma\acute{\gamma}}\zeta_{\lambda\acute{\lambda}}
\simeq{1\over 2}(\eta_{\gamma\lambda}\eta_{\acute{\gamma}\acute{\lambda}}
-\eta_{\gamma\acute{\lambda}}\eta_{\acute{\gamma}\lambda}).
$
The symbol $\simeq$ means that the equation holds up to longitudinal terms.
The second-rank tensor gauge field $A_{\alpha\acute{\alpha}}$ with 16
components  describes in this theory three
physical transversal polarizations.
The propagator $\Delta_{\gamma\acute{\gamma}\lambda\acute{\lambda}}(k)$
is defined through the equation
$
H^{fix}_{\alpha\acute{\alpha}\gamma\acute{\gamma}}(k)
\Delta^{\gamma\acute{\gamma}}_{~~~\lambda\acute{\lambda}}(k) =
 \eta_{\alpha\lambda}\eta_{\acute{\alpha}\acute{\lambda}}~~,
$
and has the following form:
\be
\Delta_{\gamma\acute{\gamma}\lambda\acute{\lambda}}(k) = -
{4 \eta_{\gamma\lambda}\eta_{ \gamma^{'} \lambda^{'}}
+2 \eta_{\gamma\lambda^{'}}\eta_{ \gamma^{'}\lambda }
-3\eta_{\gamma \gamma^{'}}\eta_{\lambda \lambda^{'}}
 \over 3(k^2 - i\varepsilon)}~~.
\ee
The corresponding residue can be represented as a sum
\beqa
{4 \eta_{\gamma\lambda}\eta_{ \gamma^{'} \lambda^{'}}
+2 \eta_{\gamma\lambda^{'}}\eta_{ \gamma^{'}\lambda }
-3\eta_{\gamma \gamma^{'}}\eta_{\lambda \lambda^{'}}
 \over 3 }
=&& (\eta_{\gamma\lambda}\eta_{\acute{\gamma}\acute{\lambda}}
+\eta_{\gamma\acute{\lambda}}\eta_{\acute{\gamma}\lambda}
-\eta_{\gamma\acute{\gamma}}\eta_{\lambda\acute{\lambda}})+
 {1\over 3}(\eta_{\gamma\lambda}\eta_{\acute{\gamma}\acute{\lambda}}
-\eta_{\gamma\acute{\lambda}}\eta_{\acute{\gamma}\lambda}).\nn
\eeqa
The first term describes the $\lambda= \pm 2$ helicity states and is
represented by the symmetric part $\varepsilon_{\alpha\acute{\alpha }}$ of the polarization tensor,
the second term describes $\lambda= 0$ helicity state and is represented
by its antisymmetric part $\zeta_{\alpha\acute{\alpha}}$.

Let us now consider three-particle interaction vertex - VTT.
Explicitly first three-linear term of the Lagrangian (\ref{Lagrangian})
has the following form:
\beqa\label{cubicterm}
&-&{1 \over 2}g f^{abc}(\partial_{\mu} A^{a}_{\nu\lambda} -
\partial_{\nu} A^{a}_{\mu\lambda})~ (A^{b}_{\mu}A^{c}_{\nu\lambda}+
A^{b}_{\mu\lambda}A^{c}_{\nu})
-{1 \over 4}g f^{abc}(\partial_{\mu} A^{a}_{\nu} -
\partial_{\nu} A^{a}_{\mu})~ 2A^{b}_{\mu\lambda}A^{c}_{\nu\lambda}.\nn
\eeqa
This is in addition to the standard Yang-Mills VVV three-vector boson interaction vertex
$$
{{\cal L}}^{cubic}_{1} = -{1 \over 2} g f^{abc}(\partial_{\mu} A^{a}_{\nu} -
\partial_{\nu} A^{a}_{\mu})
A^{b}_{\mu} A^{c}_{\nu},
$$
which in the momentum representation has the form
\be\label{VVV}
{{\cal V}}^{abc}_{\alpha\beta\gamma}(k,p,q)= -i g f^{abc}
 F_{\alpha\beta\gamma}(k,p,q) =
-i g f^{abc} [\eta_{\alpha\beta} (p-k)_{\gamma}+ \eta_{\alpha\gamma} (k-q)_{\beta}
 + \eta_{\beta\gamma} (q-p)_{\alpha}] .
\ee
In momentum space our vertex has the form
\be
-  i g f^{abc} F_{\alpha\acute{\alpha}\beta\gamma\acute{\gamma}}(k,p,q)
=-  i g f^{abc} [\eta_{\alpha\beta} (p-k)_{\gamma}+ \eta_{\alpha\gamma} (k-q)_{\beta}
 + \eta_{\beta\gamma} (q-p)_{\alpha}] \eta_{\acute{\alpha}\acute{\gamma}}.
\ee
\begin{figure}
\centerline{\hbox{\psfig{figure=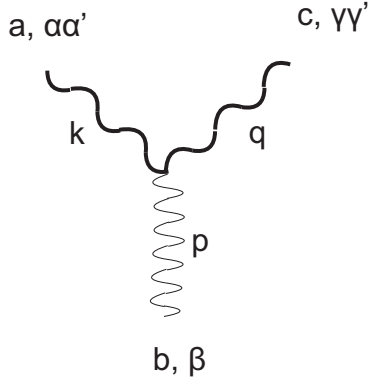,height=5cm,angle=0}}}
\caption[fig2]{The interaction vertex for the vector gauge boson V and two
tensor gauge bosons T - the VTT vertex -
$\CV^{abc}_{\alpha\acute{\alpha}\beta\gamma\acute{\gamma}}(k,p,q)$
 in non-Abelian tensor gauge
field theory \cite{Savvidy:2005ki}.
Vector gauge bosons are conventionally drawn
as thin wave lines, tensor gauge bosons are thick wave lines.
The Lorentz indices $\alpha\acute{\alpha}$ and momentum $k$ belong to the
first tensor gauge boson, the $\gamma\acute{\gamma}$ and momentum $q$
belong to the second tensor gauge boson, and Lorentz index $\beta$  and
momentum $p$ belong to the vector gauge boson. }
\label{fig2}
\end{figure}
We have also a second part of the three-particle interaction vertex VTT.
Explicitly the second
three-linear term of Lagrangian (\ref{Lagrangian}) has the following form:
\beqa
&+&{1\over 2}g f^{abc}(\partial_{\mu} A^{a}_{\nu\lambda} -
\partial_{\nu} A^{a}_{\mu\lambda})~ (A^{b}_{\mu}A^{c}_{\lambda\nu}+
A^{b}_{\mu\nu}A^{c}_{\lambda})\nonumber\\
&+&{1\over 2}g f^{abc}(\partial_{\mu} A^{a}_{\nu\nu} -
\partial_{\nu} A^{a}_{\mu\nu})~ (A^{b}_{\mu}A^{c}_{\lambda\lambda}+
A^{b}_{\mu\lambda}A^{c}_{\lambda})\nonumber\\
&+&{1\over 2}g f^{abc}(\partial_{\mu} A^{a}_{\nu} -
\partial_{\nu} A^{a}_{\mu})~ (A^{b}_{\mu\nu}A^{c}_{\lambda\lambda}+
A^{b}_{\mu\lambda}A^{c}_{\lambda\nu}) ,
\eeqa
so that in the momentum space we have
\beqa
i g  {1\over 2}f^{abc} F^{'}_{\alpha\acute{\alpha}\beta\gamma\acute{\gamma}}(k,p,q) =
i g  {1\over 2}f^{abc}[&+&(p-k)_{\gamma}(\eta_{\alpha\acute{\gamma}}
\eta_{\acute{\alpha}\beta}+
\eta_{\alpha\acute{\alpha}} \eta_{\beta\acute{\gamma}})\nn\\
&+& (k-q)_{\beta}(\eta_{\alpha\acute{\gamma}} \eta_{\acute{\alpha}\gamma}+
\eta_{\alpha\acute{\alpha}} \eta_{\gamma\acute{\gamma}})\nn\\
&+& (q-p)_{\alpha} (\eta_{\acute{\alpha}\gamma} \eta_{\beta\acute{\gamma}}+
\eta_{\acute{\alpha}\beta} \eta_{\gamma\acute{\gamma}})\nn\\
&+&(p-k)_{\acute{\alpha}}\eta_{\alpha\beta} \eta_{\gamma\acute{\gamma}}+
(p-k)_{\acute{\gamma}} \eta_{\alpha\beta} \eta_{\acute{\alpha}\gamma}\nn\\
&+&(k-q)_{\acute{\alpha}} \eta_{\alpha\gamma} \eta_{\beta\acute{\gamma}}+
(k-q)_{\acute{\gamma}}\eta_{\alpha\gamma} \eta_{\acute{\alpha}\beta}\nn\\
&+&(q-p)_{\acute{\alpha}} \eta_{\beta\gamma} \eta_{\alpha\acute{\gamma}}+
(q-p)_{\acute{\gamma}}\eta_{\alpha\acute{\alpha}} \eta_{\beta\gamma}].
\eeqa
Collecting two terms of the three-point vertex VTT together we shall get
\cite{Savvidy:2005ki}
\be\label{VTT}
\CV^{abc}_{\alpha\acute{\alpha}\beta\gamma\acute{\gamma}}(k,p,q) =
-  i g f^{abc}\{
F_{\alpha\acute{\alpha}\beta\gamma\acute{\gamma}} -
 {1\over 2}F^{'}_{\alpha\acute{\alpha}\beta\gamma\acute{\gamma}} \} \equiv
-  i g f^{abc}
\CF_{\alpha\acute{\alpha}\beta\gamma\acute{\gamma}}
\ee
where the indices $(a,\alpha,\acute{\alpha},k)$ belongs to the tensor
gauge boson, $(b,\beta,p)$ to the vector gauge boson and $(c,\gamma,\acute{\gamma},q)$
to the second tensor gauge boson (see Fig.{\ref{fig2}).

Let us consider now four-particle interaction terms of the Lagrangian
(\ref{totalactiontwo}). We have the
standard  four vector boson interaction vertex VVVV
\beqa\label{VVVV}
{{\cal V}}^{abcd}_{\alpha\beta\gamma\delta}(k,p,q,r) = - g^2 f^{lac}f^{lbd} (\eta_{\alpha \beta}
\eta_{\gamma\delta} - \eta_{\alpha \delta} \eta_{\beta\gamma})\nonumber\\
-g^2 f^{lad}f^{lbc} (\eta_{\alpha \beta} \eta_{\gamma\delta} -
\eta_{\alpha \gamma}\eta_{\beta\delta} )\nonumber\\
-g^2 f^{lab}f^{lcd} (\eta_{\alpha \gamma} \eta_{\beta\delta} -
\eta_{\alpha \delta}\eta_{\beta\gamma} )
\eeqa
and a new interaction of two vector and two tensor gauge bosons - the VVTT vertex,
\beqa
&-&{1 \over 4}g^2 f^{abc}f^{a\acute{b}\acute{c}}
(A^{b}_{\mu} A^{c}_{\nu\lambda} +
A^{b}_{\mu\lambda} A^{c}_{\nu})(
A^{\acute{b}}_{\mu}A^{\acute{c}}_{\nu\lambda} +
~A^{\acute{b}}_{\mu\lambda}A^{\acute{c}}_{\nu})\nonumber\\
&-&{1 \over 2}g^2 f^{abc}f^{a\acute{b}\acute{c}}
A^{b}_{\mu} A^{c}_{\nu}A^{\acute{b}}_{\mu\lambda}A^{\acute{c}}_{\nu\lambda},
\eeqa
which in the momentum space will take the following form:
\begin{figure}
\centerline{\hbox{\psfig{figure=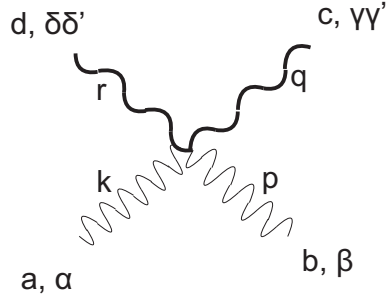,width=5cm}}}
\caption[fig3]{The quartic vertex with two vector gauge bosons and two
tensor gauge bosons - the VVTT vertex -
${{\cal V}}^{abcd}_{\alpha\beta\gamma\acute{\gamma}\delta\acute{\delta}}(k,p,q,r)$
in non-Abelian tensor gauge
field theory \cite{Savvidy:2005ki}.
Vector gauge bosons are conventionally drawn
as thin wave lines, tensor gauge bosons are thick wave lines.
The Lorentz indices $\gamma\acute{\gamma}$ and momentum $q$ belong to the
first tensor gauge boson, $\delta\acute{\delta}$ and momentum $r$ belong to the
second tensor gauge boson, the index $\alpha$  and momentum $k$
belong to the first vector gauge boson and Lorentz index $\beta$  and
momentum $p$ belong to the second vector gauge boson.
}
\label{fig3}
\end{figure}
\beqa
F^{abcd}_{\alpha\beta\gamma\acute{\gamma}\delta\acute{\delta}}(k,p,q,r)=
&-&  g^2 f^{lac}f^{lbd} (\eta_{\alpha \beta}
\eta_{\gamma\delta} - \eta_{\alpha \delta} \eta_{\beta\gamma})
\eta_{\acute{\gamma}\acute{\delta}}\nonumber\\
&-& g^2 f^{lad}f^{lbc} (\eta_{\alpha \beta} \eta_{\gamma\delta} -
\eta_{\alpha \gamma}\eta_{\beta\delta} )\eta_{\acute{\gamma}\acute{\delta}}\nonumber\\
&-& g^2 f^{lab}f^{lcd} (\eta_{\alpha \gamma} \eta_{\beta\delta} -
\eta_{\alpha \delta}\eta_{\beta\gamma} )\eta_{\acute{\gamma}\acute{\delta}}.
\eeqa
The second part of the vertex VVTT is:
\beqa
&+&{1 \over 4}g^2 f^{abc}f^{a\acute{b}\acute{c}}
(A^{b}_{\mu} A^{c}_{\nu\lambda} +
A^{b}_{\mu\lambda} A^{c}_{\nu})
(A^{\acute{b}}_{\mu}A^{\acute{c}}_{\lambda\nu} +
~A^{\acute{b}}_{\mu\nu}A^{\acute{c}}_{\lambda})\nonumber\\
&+&{1 \over 4}g^2 f^{abc}f^{a\acute{b}\acute{c}}
(A^{b}_{\mu} A^{c}_{\nu\nu} + A^{b}_{\mu\nu} A^{c}_{\nu})
(A^{\acute{b}}_{\mu}A^{\acute{c}}_{\lambda\lambda} +
~A^{\acute{b}}_{\mu\lambda}A^{\acute{c}}_{\lambda})\nonumber\\
&+&{1 \over 2}g^2 f^{abc}f^{a\acute{b}\acute{c}}
A^{b}_{\mu} A^{c}_{\nu}
(A^{\acute{b}}_{\mu\nu}A^{\acute{c}}_{\lambda\lambda} +
A^{\acute{b}}_{\mu\lambda}A^{\acute{c}}_{\lambda\nu}),
\eeqa
which in the momentum representation will take the form
\beqa
\acute{F}^{abcd}_{\alpha\beta\gamma\acute{\gamma}\delta\acute{\delta}}(k,p,q,r)=
{1 \over 2}g^2 f^{lac}f^{lbd} [&+&\eta_{\alpha \beta}
(\eta_{\gamma\acute{\delta}}\eta_{\acute{\gamma}\delta} +
\eta_{\gamma\acute{\gamma}} \eta_{\delta\acute{\delta}})\nn\\
&-&\eta_{\beta\gamma}
(\eta_{\alpha \acute{\delta}}\eta_{\acute{\gamma}\delta} +
\eta_{\alpha\acute{\gamma}} \eta_{\delta\acute{\delta}})\nn\\
&-&\eta_{\alpha\delta}
(\eta_{\beta\acute{\gamma}}\eta_{\gamma\acute{\delta}} +
\eta_{\beta\acute{\delta}} \eta_{\gamma\acute{\gamma}})\nn\\
&+&\eta_{\gamma\delta}
(\eta_{\alpha\acute{\delta}}\eta_{\beta\acute{\gamma}} +
\eta_{\alpha\acute{\gamma}} \eta_{\beta\acute{\delta}})]\nn\\
{1 \over 2}g^2 f^{lad}f^{lbc} [&+&\eta_{\alpha \beta}
(\eta_{\gamma\acute{\delta}}\eta_{\acute{\gamma}\delta} +
\eta_{\gamma\acute{\gamma}} \eta_{\delta\acute{\delta}})\nn\\
&-&\eta_{\alpha\gamma}
(\eta_{\beta\acute{\delta}}\eta_{\acute{\gamma}\delta} +
\eta_{\beta\acute{\gamma}} \eta_{\delta\acute{\delta}})\nn\\
&-&\eta_{\beta\delta}
(\eta_{\alpha\acute{\gamma}}\eta_{\gamma\acute{\delta}} +
\eta_{\alpha\acute{\delta}} \eta_{\gamma\acute{\gamma}})\nn\\
&+&\eta_{\gamma\delta}
(\eta_{\alpha\acute{\gamma}}\eta_{\beta\acute{\delta}} +
\eta_{\alpha\acute{\delta}} \eta_{\beta\acute{\gamma}})]\nn\\
{1 \over 2}g^2 f^{lab}f^{lcd}[&+&\eta_{\alpha \gamma}
(\eta_{\beta\acute{\gamma}}\eta_{\delta\acute{\delta}} +
\eta_{\beta\acute{\delta}} \eta_{\delta\acute{\gamma}})\nn\\
&-&\eta_{\beta\gamma}
(\eta_{\alpha\acute{\gamma}}\eta_{\delta\acute{\delta}} +
\eta_{\alpha\acute{\delta}} \eta_{\delta\acute{\gamma}})\nn\\
&-&\eta_{\alpha\delta}
(\eta_{\beta\acute{\delta}}\eta_{\gamma\acute{\gamma}} +
\eta_{\beta\acute{\gamma}} \eta_{\gamma\acute{\delta}})\nn\\
&+&\eta_{\beta\delta}
(\eta_{\alpha\acute{\delta}}\eta_{\gamma\acute{\gamma}} +
\eta_{\alpha\acute{\gamma}} \eta_{\gamma\acute{\delta}})].
\eeqa
The total vertex is
\be\label{VVTT}
{{\cal V}}^{abcd}_{\alpha\beta\gamma\acute{\gamma}\delta\acute{\delta}}(k,p,q,r)=
F^{abcd}_{\alpha\beta\gamma\acute{\gamma}\delta\acute{\delta}}(k,p,q,r) +
\acute{F}^{abcd}_{\alpha\beta\gamma\acute{\gamma}\delta\acute{\delta}}(k,p,q,r).
\ee
In summary we have off-mass-shell  Yang-Mills vertex VVV (\ref{VVV}), the new
vertex  VTT (\ref{VTT}) together with Yang-Mills vertex VVVV (\ref{VVVV})
and the new vertex VVTT (\ref{VVTT}) (see Fig.{\ref{fig2},{\ref{fig3}).

\subsection{\it Three-Point Amlitudes }

In order to compare these vertices with the corresponding tree-level amplitudes of the
open string theory one should project them to the mass-shell,
because  the string amplitudes can be computed only on the mass-shell
\footnote{We shall derive the corresponding string amplitudes in the next section.}.
The three point scattering amplitudes for massless particles
are equal to zero for the real momenta $(k,p,q)$, but if one allows complex momenta or a
different space-time signature
\cite{Berends:1981rb,Kleiss:1985yh,Xu:1986xb,Gunion:1985vca,Dixon:1996wi,Parke:1986gb,Berends:1987me,
Witten:2003nn,Britto:2004ap,Britto:2005fq,Benincasa:2007xk,Bengtsson:1983pd,Bengtsson:1983pg,Bengtsson:1986kh},
then these matrix elements have nontrivial
behavior and we will be able to compare them with the open string tree-level amplitudes.

Thus multiplying the above VTT vertex (\ref{VTT}) by the vector wave function
$e^{\beta}(p)$
and tensor wave functions $f^{\alpha\acute{\alpha}}(k)$ and
$f^{\gamma\acute{\gamma}}(q)$ we shall get the amplitude $\CV f e f$:
\beqa\label{totalvertexmassshell}
\CV_{abc}^{\alpha\acute{\alpha}\beta\gamma\acute{\gamma}}(k,p,q)
\vert_{mass-shell}
= -  i g f^{abc}
[&+(k-q)^{\beta}  (\eta^{\alpha\gamma}\eta^{\alpha^{'}\gamma^{'}} - {1\over 2}
\eta^{\alpha\gamma^{'}} \eta^{\alpha^{'}\gamma} ) \nn\\
&+(q-p )^{\alpha} (\eta^{\beta\gamma}\eta^{\alpha^{'}\gamma^{'}} - {1\over 2}
\eta^{\beta\gamma^{'}} \eta^{\alpha^{'}\gamma} )  \nn\\
&-{1\over 2}(q -p)^{\alpha^{'}} (\eta^{\beta\gamma}\eta^{\alpha \gamma^{'}} - {1\over 2}
\eta^{\beta\gamma^{'}} \eta^{\alpha \gamma} ) \nn\\
&+(p-k)^{\gamma} (\eta^{\alpha\beta}\eta^{\alpha^{'} \gamma^{'}} - {1\over 2}
\eta^{\alpha^{'}\beta} \eta^{\alpha \gamma^{'}} ) \nn\\
&-{1\over 2}(p-k)^{\gamma^{'}} (\eta^{\alpha\beta}\eta^{\alpha^{'} \gamma } - {1\over 2}
\eta^{\alpha^{'}\beta} \eta^{\alpha \gamma} ) ].
\eeqa
Here we have used the transversality of the wave functions (\ref{wavefunctions}):
\be
p_{\beta} e^{\beta}(p)=0,~~~
k_{\alpha}f^{\alpha\acute{\alpha}}(k)=k_{\acute{\alpha}}f^{\alpha\acute{\alpha}}(k)=0,~~~
q_{\gamma}f^{\gamma\acute{\gamma}}(q)=q_{\acute{\gamma}}f^{\gamma\acute{\gamma}}(k)=0.
\ee
and that they are traceless $tr f(k) = tr f(q)=0$. In (\ref{totalvertexmassshell})
and in the subsequent formulas for amplitudes we shall not show the wave functions. The tensor wave function
$f_{\alpha\acute{\alpha}}$ here is a sum of symmetric
$\varepsilon_{\alpha\acute{\alpha}}$ and antisymmetric
$\zeta_{\alpha\acute{\alpha}}$ parts (\ref{wavefunctions}).

We shell separate the parts of this vertex which are symmetric - $T_S$ and
antisymmetric -$T_A$ with respect to the indices of the tensor field of
$f_{\alpha\acute{\alpha}}= \varepsilon_{\alpha\acute{\alpha}}
 + \zeta_{\alpha\acute{\alpha}}$ .
The symmetric $\CV \varepsilon e \varepsilon$ part of the amplitude  is
\beqa\label{ymsymmetricvertex}
\CV_{abc}^{\alpha\acute{\alpha}\beta\gamma\acute{\gamma}}(k,p,q)
\vert_{mass-shell}
= -  i{1\over 4}g f^{abc}
[&+(k-q)^{\beta}  (\eta^{\alpha\gamma}\eta^{\alpha^{'}\gamma^{'}} +
\eta^{\alpha\gamma^{'}} \eta^{\alpha^{'}\gamma} ) \nn\\
&+{1\over 4}(q-p )^{\alpha} (\eta^{\beta\gamma}\eta^{\alpha^{'}\gamma^{'}} +
\eta^{\beta\gamma^{'}} \eta^{\alpha^{'}\gamma} )  \nn\\
&+{1\over 4}(q- p)^{\alpha^{'}} (\eta^{\beta\gamma}\eta^{\alpha \gamma^{'}} +
\eta^{\beta\gamma^{'}} \eta^{\alpha \gamma} ) \nn\\
&+{1\over 4}(p-k)^{\gamma} (\eta^{\alpha\beta}\eta^{\alpha^{'} \gamma^{'}} +
\eta^{\alpha^{'}\beta} \eta^{\alpha \gamma^{'}} ) \nn\\
&+{1\over 4}(p-k)^{\gamma^{'}} (\eta^{\alpha\beta}\eta^{\alpha^{'} \gamma } +
\eta^{\alpha^{'}\beta} \eta^{\alpha \gamma} ) ].
\eeqa
The antisymmetric $\CV \zeta e \zeta$ part of the amplitude is
\beqa\label{ymantisymmetricvertex}
\CV_{abc}^{\alpha\acute{\alpha}\beta\gamma\acute{\gamma}}(k,p,q)
\vert_{mass-shell}
= -  i {3\over 4} g f^{abc}
[&+(k-q)^{\beta}  (\eta^{\alpha\gamma}\eta^{\alpha^{'}\gamma^{'}} -
\eta^{\alpha\gamma^{'}} \eta^{\alpha^{'}\gamma} ) \nn\\
&+{3\over 4}(q-p )^{\alpha} (\eta^{\beta\gamma}\eta^{\alpha^{'}\gamma^{'}} -
\eta^{\beta\gamma^{'}} \eta^{\alpha^{'}\gamma} )  \nn\\
&-{3\over 4}(q- p)^{\alpha^{'}} (\eta^{\beta\gamma}\eta^{\alpha \gamma^{'}}-
\eta^{\beta\gamma^{'}} \eta^{\alpha \gamma} ) \nn\\
&+{3\over 4}(p-k)^{\gamma} (\eta^{\alpha\beta}\eta^{\alpha^{'} \gamma^{'}} -
\eta^{\alpha^{'}\beta} \eta^{\alpha \gamma^{'}} ) \nn\\
&-{3\over 4}(p-k)^{\gamma^{'}} (\eta^{\alpha\beta}\eta^{\alpha^{'} \gamma } -
\eta^{\alpha^{'}\beta} \eta^{\alpha \gamma} ) ].
\eeqa
The mixed symmetry  part $\CV \varepsilon e \zeta$ of the amplitude is
\beqa\label{mixedsymmetryvertex1}
\CV_{abc}^{\alpha\acute{\alpha}\beta\gamma\acute{\gamma}}(k,p,q)
\vert_{mass-shell}
= -  i {3\over 16} g f^{abc}
[&+(q-p )^{\alpha} (\eta^{\alpha^{'}\gamma^{'}}\eta^{\beta\gamma} -
\eta^{\alpha^{'}\gamma}\eta^{\beta\gamma^{'}}  )  \nn\\
&+(q- p)^{\alpha^{'}} (\eta^{\alpha \gamma^{'}} \eta^{\beta\gamma}-
\eta^{\alpha \gamma}\eta^{\beta\gamma^{'}}) \nn\\
&+(p-k)^{\gamma} (\eta^{\alpha\beta}\eta^{\alpha^{'} \gamma^{'}} +
\eta^{\alpha \gamma^{'}}\eta^{\alpha^{'}\beta}  ) \nn\\
&-(p-k)^{\gamma^{'}} (\eta^{\alpha\beta}\eta^{\alpha^{'} \gamma } +
\eta^{\alpha \gamma} \eta^{\alpha^{'}\beta}  ) ].
\eeqa
The last vertex is symmetric under the interchange
($\alpha  \leftrightarrow \alpha^{'}$) and
antisymmetric under ($\gamma \leftrightarrow \gamma^{'}$).
There is also mixed symmetry part of the vertex which is antisymmetric in
($\alpha  \leftrightarrow \alpha^{'}$) and
symmetric under ($\gamma \leftrightarrow \gamma^{'}$) the $\CV \zeta e \varepsilon $ amplitude:
\beqa\label{mixedsymmetryvertex2}
\CV_{abc}^{\alpha\acute{\alpha}\beta\gamma\acute{\gamma}}(k,p,q)
\vert_{mass-shell}
= -  i {3\over 16} g f^{abc}
[&+(q-p )^{\alpha} (\eta^{\alpha^{'}\gamma^{'}}\eta^{\beta\gamma} +
\eta^{\alpha^{'}\gamma}\eta^{\beta\gamma^{'}}  )  \nn\\
&-(q- p)^{\alpha^{'}} (\eta^{\alpha \gamma^{'}} \eta^{\beta\gamma}+
\eta^{\alpha \gamma}\eta^{\beta\gamma^{'}}) \nn\\
&+(p-k)^{\gamma} (\eta^{\alpha\beta}\eta^{\alpha^{'} \gamma^{'}} -
\eta^{\alpha \gamma^{'}}\eta^{\alpha^{'}\beta}  ) \nn\\
&+(p-k)^{\gamma^{'}} (\eta^{\alpha\beta}\eta^{\alpha^{'} \gamma } -
\eta^{\alpha \gamma} \eta^{\alpha^{'}\beta}  ) ]
\eeqa
One can check that the sum of the above four terms (\ref{ymsymmetricvertex}),
(\ref{ymantisymmetricvertex}),(\ref{mixedsymmetryvertex1}) and (\ref{mixedsymmetryvertex2}) gives the total
vertex (\ref{totalvertexmassshell}).

\subsection{\it Complex Deformation of Momenta}

The  nontrivial three-point amplitudes can be defined if one considers
complex momenta or the space-time signature $\eta^{\mu\nu}=(-+-+)$ .
Then the momenta can be chosen
as follows \cite{Britto:2005fq}:
$$
k^{\mu}_{1}= (\omega, z, iz, k),~~k^{\mu}_{2}= (\omega, -z, -iz, k),~~k^{\mu}_{3}= (2\omega, 0, 0, 2k),
$$
to fulfill the momentum conservation
$$
k_1 +k_2 =k_3.
$$
All massless bosons are on mass-shell   $k^{2}_{1}=k^{2}_{2}=k^{2}_{3}=0,~~~(\omega^2 = k^2)$
and
$$
k_1 \cdot k_2=k_2 \cdot k_3=k_3 \cdot k_1=0.
$$
Let us first consider the matrix element VVV for the vector gauge bosons. The polarization vectors are
$$
e^{+}_{1}={1\over \sqrt{2}} ({z \over \omega}, 1, -i, -{z \over k} ),~~
e^{+}_{2}={1\over \sqrt{2}}(-{z \over \omega}, 1, -i, {z \over k} ),~~
e^{-}_{3}={1\over \sqrt{2}}(0, 1,  i ,0 )
$$
and are orthogonal to the corresponding momenta
$$
k_1 \cdot e^{+}_{1} =0,~~k_2 \cdot e^{+}_{2} =0,~~k_3 \cdot e^{-}_{3} =0.
$$
We can compute now the matrix element using trilinear vertex VVV (\ref{VVV}):
\beqa
&M(+,+,-)= F^{\mu_1\mu_2\mu_3} (k_1 ,k_2 ,k_3) e^+_{\mu_1}(k_1)e^+_{\mu_2}(k_2)e^-_{\mu_3}(k_3)=\nn\\
&=-2 e^+_{1} \cdot e^+_{2}~ k_{1} \cdot e^-_{3} -
2 e^+_{2} \cdot e^-_{3}~ k_{2} \cdot e^+_{1} +
2 e^-_{3} \cdot e^+_{1}~ k_{3} \cdot e^+_{2} = 8 \sqrt{2} z .
\eeqa
And indeed, it is nonzero and  grows linearly with momentum deformation z.
Using spinor representation of the momenta and polarization vectors \cite{Witten:2003nn}
$$
k_{a\dot{a}}= \lambda_a \tilde{\lambda}_{\dot{a}},~~e^{+}_{a\dot{a}}
={\mu_a \tilde{\lambda}_{\dot{a}} \over <\mu,\lambda>},~~e^{-}_{a\dot{a}}
={\lambda_a \tilde{\mu}_{\dot{a}} \over [\lambda,\mu]},
$$
where
$$
\lambda_a = (\sqrt{k^+},{k_x +i k_y \over \sqrt{k^+}}),~~~
\tilde{\lambda}_{\dot{a}} = (\sqrt{k^+},{k_x - i k_y \over \sqrt{k^+}}),~~~~k^+=k_t + k_z~,
$$
one can see that
$$
<1,2>=<2,3>=<3,1> =0,~~~
$$
but
$$
[1,2]=-4 z,~~~[2,3]= 2\sqrt{2}z,~~~[3,1]= 2\sqrt{2}z,
$$
and
$$
M(+,+,-)=- \sqrt{2}~  {[1,2]^4 \over [1,2] [2,3] [3,1]}.
$$
We can calculate now the trilinear vertex VTT matrix element using expressions
(\ref{VTT}), (\ref{ymsymmetricvertex})
\beqa\label{ymantisymmetricvertexspinorial}
&M(+2,+1,-2)= \CF^{\alpha\acute{\alpha}\beta\gamma\acute{\gamma}}(k_1 ,k_2 ,k_3)
\varepsilon^+_{\alpha\acute{\alpha}}(k_1)e^+_{\beta}(k_2)\varepsilon^-_{\gamma\acute{\gamma}}(k_3)
=\nn\\
&=2 (k_{1}+k_{3}) \cdot e^+_{2} ~\varepsilon^+_{1} \cdot \varepsilon^-_{3}~  +
 (-k_3- k_{2}) \cdot \varepsilon^+_{1} \cdot \varepsilon^-_{3} \cdot e^+_{2}~+
(k_2-k_{1}) \cdot \varepsilon^-_{3} \cdot \varepsilon^+_{1} \cdot e^+_{2}~ \nn\\
&= 12 \sqrt{2} z,
\eeqa
where $\varepsilon^+_{\alpha\acute{\alpha}}(k_1)= e^{+}_{\alpha}(k_1) e^{+}_{\acute{\alpha}}(k_1)$,~
$\varepsilon^-_{\gamma\acute{\gamma}}(k_3)= e^{-}_{\gamma}(k_3) e^{-}_{\acute{\gamma}}(k_3)$.
It is nonzero and  also grows linearly with momentum deformation z.
Using spinor representation we shall get
\be
M(+2,+1,-2)=  - {3\sqrt{2} \over 4}{[1,2]^6 \over [1,2] [2,3]^3 [3,1]}.
\ee

\subsection{\it  The Dual Lagrangian}

For completeness let us also recall expression for the dual Lagrangian which is defined as follows
\cite{Barrett:2007,Guttenberg:2008zn}:
\beqa\label{alternativetotalactiontwo}
\tilde{{{\cal L}}}_2   =
 &-&{1\over 4}\tilde{G}^{a}_{\mu\nu,\lambda}\tilde{G}^{a}_{\mu\nu,\lambda}
-{1\over 4}G^{a}_{\mu\nu}\tilde{G}^{a}_{\mu\nu,\lambda\lambda} +\nn\\
&+&{1\over 4}\tilde{G}^{a}_{\mu\nu,\lambda}\tilde{G}^{a}_{\mu\lambda,\nu}
+{1\over 4}\tilde{G}^{a}_{\mu\nu,\nu}\tilde{G}^{a}_{\mu\lambda,\lambda}
+{1\over 2}G^{a}_{\mu\nu}\tilde{G}^{a}_{\mu\lambda,\nu\lambda},
\eeqa
where the dual field strength tensors are:
\beqa\label{fieldstrengthtensors}
\tilde{G}^{a}_{\mu\nu,\lambda} &=&
\partial_{\mu} A^{a}_{\lambda\nu} - \partial_{\nu} A^{a}_{\lambda\mu} +
g f^{abc}(~A^{b}_{\mu}~A^{c}_{\lambda\nu} + A^{b}_{\lambda\mu}~A^{c}_{\nu} ~),
\nn\\
\tilde{G}^{a}_{\mu\nu,\lambda\rho} &=&  \{~
\partial_{\mu}({2\over 3} A^{a}_{\lambda\nu\rho}+{2\over 3}A^{a}_{\rho\nu\lambda}
-{1\over 3}A^{a}_{\nu\lambda\rho})
+ g f^{abc}~A^{b}_{\mu}~
({2\over 3} A^{a}_{\lambda\nu\rho}+{2\over 3}A^{a}_{\rho\nu\lambda} -{1\over 3}A^{a}_{\nu\lambda\rho}) +\nn\\
&-& \partial_{\nu} ( {2\over 3}A^{a}_{\lambda\mu\rho} + {2\over 3}A^{a}_{\rho\mu\lambda}- {1\over 3}A^{a}_{\mu\lambda\rho}) +
g f^{abc}~
({2\over 3} A^{a}_{\lambda\mu\rho} + {2\over 3}A^{a}_{\rho\mu\lambda}-
{1\over 3}A^{a}_{\mu\lambda\rho})~A^{c}_{\nu} ~\}
\nn\\
&+& g f^{abc}~(~A^{b}_{\lambda\mu}~A^{c}_{\rho\nu}+
A^{b}_{\rho\mu}~A^{c}_{\lambda\nu}~).
\eeqa
Here we have a similar polynomial expansion of the dual Lagrangian:
\be
\tilde{{\cal L}}_2 =  {1 \over 2} A^{a}_{\alpha\acute{\alpha}}
\tilde{H}_{\alpha\acute{\alpha}\gamma\acute{\gamma}} A^{a}_{\gamma\acute{\gamma}}+
{1 \over 1! 2!}\tilde{V}^{abc}_{ \alpha \acute{\alpha} \beta \gamma\acute{\gamma} }
A^{a}_{\alpha\acute{\alpha}} A^{b}_{\beta}A^{c}_{\gamma\acute{\gamma}}+
{1 \over 2! 2!}
\tilde{V}^{abcd}_{ \alpha \beta \gamma\acute{\gamma} \delta\acute{\delta }}
A^{a}_{\alpha } A^{b}_{\beta} A^{c}_{\gamma\acute{\gamma}}
A^{d}_{\delta\acute{\delta}} +...
\ee
The kinetic term is identical with the kinetic term (\ref{basickineticterm}) of the Lagrangian
$\CL_2$
\beqa\label{dualkineticterm}
\tilde{H}_{\alpha\acute{\alpha}\gamma\acute{\gamma}}(k)=
(-\eta_{\alpha\gamma}\eta_{\acute{\alpha}\acute{\gamma}}
+{1 \over 2}\eta_{\alpha\acute{\gamma}}\eta_{\acute{\alpha}\gamma}
+{1 \over 2}\eta_{\alpha\acute{\alpha}}\eta_{\gamma\acute{\gamma}})k^2
+\eta_{\alpha\gamma}k_{\acute\alpha}k_{\acute{\gamma}}
+\eta_{\acute\alpha \acute{\gamma}}k_{\alpha}k_{\gamma}\nn\\
-{1 \over 2}(\eta_{\alpha\acute{\gamma}}k_{\acute\alpha}k_{\gamma}
+\eta_{\acute\alpha\gamma}k_{\alpha}k_{\acute{\gamma}}
+\eta_{\alpha\acute\alpha}k_{\gamma}k_{\acute{\gamma}}
+\eta_{\gamma\acute{\gamma}}k_{\alpha}k_{\acute\alpha})
\eeqa
and is fully gauge invariant operator
$
k_{\alpha}\tilde{H}_{\alpha\acute{\alpha}\gamma\acute{\gamma}}=0,~
k_{\acute{\alpha}}\tilde{H}_{\alpha\acute{\alpha}\gamma\acute{\gamma}}=0.
$
One can get also the explicit form of the dual cubic vertex  VTT  from
the dual Lagrangian $\tilde{\CL}_2$  (\ref{alternativetotalactiontwo})
\beqa\label{dualTVTveterx}
\tilde{\CV}^{abc}_{\alpha\acute{\alpha}\beta\gamma\acute{\gamma}}=
-  i g f^{abc}\{ \tilde{F}_{\alpha\acute{\alpha}\beta\gamma\acute{\gamma}} -
{1\over 2} \tilde{F}^{'}_{\alpha\acute{\alpha}\beta\gamma\acute{\gamma}} \},
\eeqa
 where
\be
\tilde{F}_{\alpha\acute{\alpha}\beta\gamma\acute{\gamma}}(k,p,q)=
[\eta_{\acute{\alpha}\beta} (p-k)_{\acute{\gamma}}+
\eta_{\acute{\alpha}\acute{\gamma}} (k-q)_{\beta}
 + \eta_{\beta\acute{\gamma}} (q-p)_{\acute{\alpha}}] \eta_{ \alpha \gamma}
\ee
and
\beqa
\tilde{F}^{'}_{\alpha\acute{\alpha}\beta\gamma\acute{\gamma}}(k,p,q) &=&
(p-k)_{\acute{\gamma}}(\eta_{\acute{\alpha} \gamma }
\eta_{ \alpha \beta}+
\eta_{\alpha\acute{\alpha}} \eta_{\beta  \gamma })\nn\\
&+& (k-q)_{\beta}(\eta_{\alpha\acute{\gamma}} \eta_{\acute{\alpha}\gamma}+
\eta_{\alpha\acute{\alpha}} \eta_{\gamma\acute{\gamma}})\nn\\
&+& (q-p)_{\acute{\alpha}} (\eta_{\alpha\acute{\gamma}} \eta_{\beta\gamma}+
\eta_{\alpha\beta} \eta_{\gamma\acute{\gamma}})\nn\\
&+&(p-k)_{\alpha}\eta_{\acute{\alpha}\beta} \eta_{\gamma\acute{\gamma}}+
(p-k)_{\gamma} \eta_{\acute{\alpha}\beta} \eta_{\alpha\acute{\gamma}}\nn\\
&+&(k-q)_{\alpha} \eta_{\acute{\alpha}\acute{\gamma}} \eta_{\beta\gamma}+
(k-q)_{\gamma}\eta_{\acute{\alpha}\acute{\gamma}} \eta_{\alpha\beta}\nn\\
&+&(q-p)_{\alpha} \eta_{\beta\acute{\gamma}} \eta_{\acute{\alpha}\gamma}+
(q-p)_{\gamma}\eta_{\alpha\acute{\alpha}} \eta_{\beta\acute{\gamma}}.
\eeqa
There is an important property of the dual vertex (\ref{dualTVTveterx})
which follows from the
fact that the above Lagrangians  (\ref{totalactiontwo}) and
(\ref{alternativetotalactiontwo}) are dual
to each other in the sense of the transformation
\beqa\label{dualityfieldtransformations}
  \begin{array}{ll}
\tilde{A}_{\alpha\acute{\alpha}} =  A_{\acute{\alpha}\alpha}   ,  \\
\tilde{A}_{\alpha\acute{\alpha}\acute{\acute{\alpha}}} =
{2\over 3}(A_{\acute{\alpha}\alpha\acute{\acute{\alpha}}} + A_{\acute{\acute{\alpha}}\alpha\acute{\alpha}})
-{1\over 3} A_{\alpha\acute{\acute{\alpha}}}.
\end{array}
\eeqa
Indeed, if one
simultaneously interchanges the indices
$\alpha \leftrightarrow \acute{\alpha}$,
$\gamma \leftrightarrow \acute{\gamma}$ of the dual cubic vertex
(\ref{dualTVTveterx}), then one can see that it will transform into the
cubic vertex (\ref{VTT}) and via versa:
\beqa
\tilde{\CV}^{abc}_{\alpha\acute{\alpha}\beta\gamma\acute{\gamma}}=
-  i g f^{abc}\{ \tilde{F}_{\alpha\acute{\alpha}\beta\gamma\acute{\gamma}} -
{1\over 2} \tilde{F}^{'}_{\alpha\acute{\alpha}\beta\gamma\acute{\gamma}}\}
 = -  i g f^{abc}\{
 F_{\acute{\alpha}\alpha\beta\acute{\gamma}\gamma} -
{1\over 2} F^{'}_{\acute{\alpha}\alpha\beta\acute{\gamma}\gamma}\}
 =\CV^{abc}_{\acute{\alpha}\alpha\beta\acute{\gamma}\gamma}
\eeqa
It is also obvious from the above relation that if one considers
the {\it self-dual sum}
\be\label{sum}
{\CL}_2 + \tilde{{\CL}}_2,
\ee
then the corresponding
VTT vertex will be self-dual in the sense that under duality
transformation (\ref{dualityfieldtransformations})
$\alpha \leftrightarrow \acute{\alpha}$,
$\gamma \leftrightarrow \acute{\gamma}$ it will be mapped into itself:
\be\label{selfdualbasic}
\CV_{\alpha\acute{\alpha}\beta\gamma\acute{\gamma}}+
\tilde{\CV}_{\alpha\acute{\alpha}\beta\gamma\acute{\gamma}}~~
\rightarrow~~
\CV_{\acute{\alpha}\alpha\beta\acute{\gamma}\gamma}+
\tilde{\CV}_{\acute{\alpha}\alpha\beta\acute{\gamma}\gamma}=
\tilde{\CV}_{\alpha\acute{\alpha}\beta\gamma\acute{\gamma}}+
\CV_{\alpha\acute{\alpha}\beta\gamma\acute{\gamma}}.
\ee

\section{{\it Open Strings Tree-Level Amplitudes}}

In this section we shall calculate a tree-level scattering amplitudes
in the open string theory with Chan-Paton charges in order to compare
them with corresponding matrix elements in non-Abelian tensor gauge field
theory. We shall consider scattering amplitudes of the first excited states
of open string depicted on the Fig.\ref{fig1}, that is the charged scalar, vector and tensor
bosons.  To set up notation let us begin with the simplest example of the tree-level
scattering amplitude for the three on-shell massless vector bosons.
The vertex operator has the following form
\cite{Green:1987sp,Polchinski:1998rq}:
\be\label{vectorsvertex}
e_{\alpha}(k):\dot{X}^{\alpha}  e^{ik  X}(y):
\ee
and we shall represent the disk as the upper half-plane so that the boundary
coordinate $y$ is real $y \in [-\infty,+\infty]$. The tree amplitude can be represented in the form
\beqa
&\CV^{\mu_1\mu_2\mu_3}_{a_{1} a_{2} a_{3} } (k_1 ,k_2 ,k_3)
= F^{\mu_1\mu_2\mu_3}(k_1 ,k_2 ,k_3)
tr(\lambda^{a_{1}} \lambda^{a_{2}}\lambda^{a_{3}}) +
 F^{\mu_2\mu_1\mu_3}(k_2 ,k_1 ,k_3)
tr(\lambda^{a_{2}} \lambda^{a_{1}} \lambda^{a_{3}})\nn:
\eeqa
where the matrix element $F$ is given below
\beqa
& F^{\mu_1\mu_2\mu_3}(k_1 ,k_2 ,k_3) = \nn\\
&=\int \prod_i ~d \mu(y_i)
<:\dot{X}^{\mu_1}e^{ik_1 X}(y_1)::\dot{X}^{\mu_2}e^{ik_2 X}(y_2):
:\dot{X}^{\mu_3}e^{ik_3 X}(y_3):>  \nn\\
& =\lim_{y_1=0,y_2=1,y_3 \rightarrow \infty}
\prod_{i<j}\vert y_i -y_j \vert^{2\alpha^{'}k_ik_j}~ y^{2}_{3}~\nn\\
&\{F^{\mu_1}_{y_1}F^{\mu_2}_{y_2} F^{\mu_2}_{y_2}  - 2\alpha^{'}
[F^{\mu_1}_{y_1} {\eta^{\mu_2\mu_3}\over (y_2 -y_3)^2} +
F^{\mu_2}_{y_2} {\eta^{\mu_3\mu_1}\over (y_3 -y_1)^2}+
F^{\mu_3}_{y_3} {\eta^{\mu_1\mu_2}\over (y_1 -y_2)^2} ]\} \nn
\eeqa
and we have to sum over two orderings of the vertex operators on the disk.
The vector functions $F^{\mu}_{y}$ are given below (\ref{theFfunctions}). All bosons are on mass-shell
$$
\alpha^{'} k^{2}_{1}=\alpha^{'} k^{2}_{2}=\alpha^{'} k^{2}_{3}=0
$$
and $k_1 +k_2 +k_3 =0$. The wave functions of the vector bosons are
\be
e_{\mu_1}(k_1),~~~e_{\mu_2}(k_2),~~~e_{\mu_3}(k_3)
\ee
and are transversal to the corresponding momenta $k_{i} \cdot e(k_i)=0,~ i=1,2,3$.
The matrix element $F^{\mu_1\mu_2\mu_3}$ has the following dimensional structure:
\be
[F^{\mu_1\mu_2\mu_3}] \sim
 (\alpha^{'})^3 (k)^3 + (\alpha^{'})^2 (k)^1 .
\ee
{\it Our intension is to extract that part of the amplitude which has dimensionless
coupling constant, that is, the last term}. We shall fix the integration measure by the choice
$y_1=0,~y_2=1,~y_3 \rightarrow \infty$:
\beqa\label{theFfunctions}
& \lim_{y_1=0,~y_2=1,~y_3 \rightarrow \infty}
\prod_{i<j}\vert y_i -y_j \vert^{2\alpha^{'}k_ik_j}~\rightarrow 1\nn\\
&F^{\mu_1}_{y_1} = -2 i \alpha^{'}~ ({k^{\mu_1}_2 \over y_1-y_2}
+{k^{\mu_1}_3 \over y_1-y_3})  \rightarrow
-2 i \alpha^{'}~ ( -k^{\mu_1}_2 )
\nn\\
& F^{\mu_2}_{y_2}= -2 i \alpha^{'}~({k^{\mu_2}_1 \over y_2-y_1}
+{k^{\mu_2}_3 \over y_2-y_3})
\rightarrow
-2 i \alpha^{'}~ (+ k^{\mu_2}_1 )\nn\\
&F^{\mu_3}_{y_3}= -2 i \alpha^{'}~({k^{\mu_3}_1 \over y_3-y_1}
+{k^{\mu_3}_2 \over y_3-y_2})\rightarrow
-2 i \alpha^{'}~ (- {k^{\mu_3}_1 \over y^{2}_{3} })
\eeqa
Thus for the $F^{\mu_1\mu_2\mu_3}(k_1 ,k_2 ,k_3)
tr(\lambda^{a_{1}} \lambda^{a_{2}}\lambda^{a_{3}})$ we have
\beqa
& i(2\alpha^{'})^2
[-k^{\mu_1}_{2}\eta^{\mu_2\mu_3} +  k^{\mu_2}_{1}\eta^{\mu_3\mu_1}-
k^{\mu_3}_{1}\eta^{\mu_1\mu_2} +
2\alpha^{'} k^{\mu_1}_{2} k^{\mu_2}_{1}k^{\mu_3}_{1}]~~
tr(\lambda^{a_{1}} \lambda^{a_{2}}\lambda^{a_{3}})
\eeqa
and adding the equal term
\beqa
& i(2\alpha^{'})^2
[+k^{\mu_1}_{3}\eta^{\mu_2\mu_3} - k^{\mu_2}_{3}\eta^{\mu_3\mu_1}+
k^{\mu_3}_{2}\eta^{\mu_1\mu_2} +
2\alpha^{'} k^{\mu_1}_{2} k^{\mu_2}_{1}k^{\mu_3}_{1}]~~
tr(\lambda^{a_{1}} \lambda^{a_{2}}\lambda^{a_{3}})
\eeqa
we can get the total matrix element together with the reversed cyclic orientation
$a_1,\mu_1,k_1 \leftrightarrow a_2,\mu_2, k_2$:
\beqa
&2i(\alpha^{'})^2
[~(k_3-k_{2})^{\mu_1}\eta^{\mu_2\mu_3} +  (k_1-k_3)^{\mu_2}\eta^{\mu_3\mu_1} +(k_2-
k_1)^{\mu_3}\eta^{\mu_1\mu_2} +\nn\\
&+ 2\alpha^{'} (k_2-k_3)^{\mu_1}  k^{\mu_2}_{1}k^{\mu_3}_{1}~]~~
 tr([\lambda^{a_{1}} , \lambda^{a_{2}}]\lambda^{a_{3}}).
\eeqa
Leaving only dimensionless coupling constant we shall have the corresponding VVV amplitude:
\beqa
&\CV^{\mu_1\mu_2\mu_3}_{a_{1} a_{2} a_{3} } (k_1 ,k_2 ,k_3) = \nn\\
&=i g ~tr([\lambda^{a_{1}} , \lambda^{a_{2}}]\lambda^{a_{3}})
[~(k_3-k_{2})^{\mu_1}\eta^{\mu_2\mu_3} +  (k_1-k_3)^{\mu_2}\eta^{\mu_3\mu_1} +(k_2-
k_1)^{\mu_3}\eta^{\mu_1\mu_2}  ~]~.~
\eeqa
which coincides with the Yang-Mills vertex (\ref{VVV}) projected to the mass-shell.
In the next subsection we shall perform a similar calculation of the scattering amplitude
between two symmetric tensor bosons  $T_S$ and a vector boson $V$ in open string theory
(see Fig.\ref{fig1}) in order to compare it with
the amplitude (\ref{ymsymmetricvertex}), (\ref{ymantisymmetricvertexspinorial})
in non-Abelian tensor gauge field theory.

\subsection{\it Tree-Level Amplitude for Two Symmetric Tensors and  Vector}

The vertex operator for the symmetric $T_S$ rank-2 tensor boson on the third level is
\be\label{symmetricvertex}
\varepsilon_{\alpha\alpha^{'}}(k):\dot{X}^{\alpha}\dot{X}^{\alpha^{'}} e^{ik  X}(y):
\ee
and together with the vertex (\ref{vectorsvertex}) can be used to calculate now the scattering
amplitude between vector and two tensor bosons:
\beqa\label{vecror2tensorvertex}
&\CV^{\alpha \alpha^{'} \beta \gamma\gamma^{'}}_{a  b c } (k  ,p ,q)
= F^{\alpha \alpha^{'} \beta \gamma\gamma^{'}}  (k  ,p ,q)~
tr(\lambda^{a } \lambda^{b}\lambda^{c}) +
 F^{\gamma\gamma^{'} \beta \alpha \alpha^{'}}  (q  ,p ,k)~
tr(\lambda^{c} \lambda^{b} \lambda^{a})
\eeqa
where the wave functions of the pair of tensor gauge bosons and the vector boson are:
\be
\varepsilon_{\alpha \alpha^{'}}(k),~~~e_{ \beta}(p),~~~\varepsilon_{\gamma\gamma^{'}}(q).
\ee
We shall define for convenience
$k_1 \equiv k, k_2 \equiv q, k_3 \equiv p,$ and  $k_1 +k_2 +k_3 =0$.
The mass-shell conditions are
$$\alpha^{'} k^{2}_{1}=\alpha^{'} k^{2}_{2}= -1,~   \alpha^{'} k^{2}_{3}=0.$$
We have to calculate the correlation function:
\beqa
&F^{\alpha \alpha^{'} \beta \gamma\gamma^{'}}_S (k  ,p ,q) = \nn\\
&=\int \prod_i ~d \mu(y_i)
<:\dot{X}^{\alpha}\dot{X}^{\alpha^{'}} e^{ik  X}(y_1):
:\dot{X}^{\gamma}\dot{X}^{\gamma^{'}} e^{iq X}(y_2):
:\dot{X}^{\beta}e^{ip X}(y_3):>                              \nn\\
& =\lim_{y_1=0,y_2=1,y_3 \rightarrow \infty}
\prod_{i<j}\vert y_i -y_j \vert^{2\alpha^{'}k_ik_j}~ y^{2}_{3}~\nn\\
&\{~ (F^{ \alpha}F^{ \alpha^{'}})_{y_1} F^{\beta}_{y_3}
(F^{ \gamma}F^{ \gamma^{'}})_{y_2} -                           \nn\\
&- 2\alpha^{'}
[ F^{ \alpha}_{y_1} F^{\beta}_{y_3}
  F^{ \gamma} _{y_2}
    {\eta^{\alpha^{'}\gamma^{'} }\over (y_1 -y_2)^2} +
F^{ \alpha}_{y_1} F^{\beta}_{y_3}
   F^{\gamma^{'} } _{y_2}
      {\eta^{\alpha^{'} \gamma}\over (y_1 -y_2)^2} +
F^{\alpha^{'} }_{y_1} F^{\beta}_{y_3}
   F^{ \gamma} _{y_2}
      {\eta^{\alpha \gamma^{'} }\over (y_1 -y_2)^2} +
F^{\alpha^{'} }_{y_1} F^{\beta}_{y_3}
    F^{\gamma^{'} } _{y_2}
       {\eta^{\alpha \gamma}\over (y_1 -y_2)^2} +                 \nn\\
&+F^{ \alpha}_{y_1} F^{\alpha^{'} }_{y_1}
  F^{ \gamma} _{y_2}
    {\eta^{\beta\gamma^{'} }\over (y_3 -y_2)^2} +
F^{ \alpha}_{y_1} F^{\alpha^{'} }_{y_1}
  F^{ \gamma^{'}} _{y_2}
    {\eta^{\beta\gamma }\over (y_3 -y_2)^2} +
F^{\alpha  }_{y_1} F^{\gamma}_{y_2}
   F^{ \gamma^{'}} _{y_2}
      {\eta^{\beta\alpha^{'}   }\over (y_3 -y_1)^2} +
F^{\alpha^{'}  }_{y_1} F^{\gamma}_{y_2}
   F^{ \gamma^{'}} _{y_2}
      {\eta^{\beta\alpha   }\over (y_3 -y_1)^2} ]  +             \nn\\
&+(2\alpha^{'})^2 [ + F^{\beta}_{y_3}
     {\eta^{\alpha \gamma  }\over (y_1 -y_2)^2}
     {\eta^{\alpha^{'}\gamma^{'} }\over (y_1 -y_2)^2} +
F^{\beta}_{y_3}
     {\eta^{\alpha \gamma^{'} }\over (y_1 -y_2)^2}
     {\eta^{\alpha^{'} \gamma }\over (y_1 -y_2)^2} + ~~~~~~~~~  \nn\\
&+  F^{\alpha}_{y_1}
     {\eta^{\beta \gamma  }\over (y_3 -y_2)^2}
     {\eta^{\alpha^{'}\gamma^{'} }\over (y_1 -y_2)^2} +
F^{\alpha}_{y_1}
     {\eta^{ \beta\gamma^{'} }\over (y_3 -y_2)^2}
     {\eta^{\alpha^{'} \gamma }\over (y_1 -y_2)^2} +   \nn\\
&+  F^{\alpha^{'}}_{y_1}
     {\eta^{\beta \gamma  }\over (y_3 -y_2)^2}
     {\eta^{\alpha \gamma^{'} }\over (y_1 -y_2)^2} +
F^{\alpha^{'}}_{y_1}
     {\eta^{ \beta\gamma^{'} }\over (y_3 -y_2)^2}
     {\eta^{\alpha \gamma }\over (y_1 -y_2)^2} +   \nn\\
&+  F^{\gamma}_{y_2}
     {\eta^{\beta \alpha }\over (y_3 -y_1)^2}
     {\eta^{\alpha^{'}\gamma^{'} }\over (y_1 -y_2)^2} +
F^{\gamma}_{y_2}
     {\eta^{ \beta\alpha^{'} }\over (y_3 -y_1)^2}
     {\eta^{\alpha \gamma^{'} }\over (y_1 -y_2)^2} +   \nn\\
&+  F^{\gamma^{'}}_{y_2}
     {\eta^{\beta \alpha }\over (y_3 -y_1)^2}
     {\eta^{\alpha^{'}\gamma  }\over (y_1 -y_2)^2} +
F^{\gamma^{'}}_{y_2}
     {\eta^{ \beta\alpha^{'} }\over (y_3 -y_1)^2}
     {\eta^{\alpha \gamma }\over (y_1 -y_2)^2}]  \}.
\eeqa
The amplitude has the following dimensional structure:
\be
F^{\alpha \alpha^{'} \beta \gamma\gamma^{'}} \sim
 (\alpha^{'})^5 (k)^5 + (\alpha^{'})^4 (k)^3 + (\alpha^{'})^3 (k)^1 ~,
\ee
where the first term contains the fifth power of $\alpha^{'}$ and the fifth power of
momentum and so on. {\it We are interested only in the last term which contains the
first power of momentum and therefore has the dimensionless coupling constant}.
We shall fix the integration measure by the choice
$y_1=0,~y_2=1,~y_3 \rightarrow \infty $~:
\beqa
& \lim_{y_1=0,~y_2=1,~y_3 \rightarrow \infty}
\prod_{i<j}\vert y_i -y_j \vert^{2\alpha^{'}k_ik_j}~\rightarrow 1,\nn\\
&F^{\alpha}_{y_1} = -2 i \alpha^{'}~ ({p^{\alpha}  \over y_1-y_3}
+{q^{\alpha}  \over y_1-y_2})  \rightarrow -2 i \alpha^{'}~( - q^{ \alpha  } ) \nn\\
& F^{\gamma}_{y_2}= -2 i \alpha^{'}~({p^{\gamma}  \over y_2-y_3 }
+{k^{\gamma} \over y_2-y_1}) \rightarrow -2 i \alpha^{'}~( k^{\gamma  } ) \nn\\
&F^{\beta}_{y_3}= -2 i \alpha^{'}~({k^{\beta}  \over y_3-y_1}
+{q^{\beta}  \over y_3-y_2}) \rightarrow  -2 i \alpha^{'}~(-{k^{\beta}  \over y^2_3} )~,
\eeqa
and keeping only part of the amplitude linear in momentum we shall get:
\beqa
F^{\alpha \alpha^{'} \beta \gamma\gamma^{'}}_S (k  ,p ,q) =- i(2\alpha^{'})^3
[&-k^{\beta}  (\eta^{\alpha\gamma}\eta^{\alpha^{'}\gamma^{'}} +
\eta^{\alpha\gamma^{'}} \eta^{\alpha^{'}\gamma} ) \nn\\
&- q^{\alpha} (\eta^{\beta\gamma}\eta^{\alpha^{'}\gamma^{'}} +
\eta^{\beta\gamma^{'}} \eta^{\alpha^{'}\gamma} )  \nn\\
&- q^{\alpha^{'}} (\eta^{\beta\gamma}\eta^{\alpha \gamma^{'}} +
\eta^{\beta\gamma^{'}} \eta^{\alpha \gamma} ) \nn\\
&+k^{\gamma} (\eta^{\alpha\beta}\eta^{\alpha^{'} \gamma^{'}} +
\eta^{\alpha^{'}\beta} \eta^{\alpha \gamma^{'}} ) \nn\\
&+k^{\gamma^{'}} (\eta^{\alpha\beta}\eta^{\alpha^{'} \gamma } +
\eta^{\alpha^{'}\beta} \eta^{\alpha \gamma} ) ].
\eeqa
We can add an equal term
\beqa
F^{\alpha \alpha^{'} \beta \gamma\gamma^{'}}_S (k  ,p ,q) =- i(2\alpha^{'})^3
[&+q^{\beta}  (\eta^{\alpha\gamma}\eta^{\alpha^{'}\gamma^{'}} +
\eta^{\alpha\gamma^{'}} \eta^{\alpha^{'}\gamma} ) \nn\\
&+p^{\alpha} (\eta^{\beta\gamma}\eta^{\alpha^{'}\gamma^{'}} +
\eta^{\beta\gamma^{'}} \eta^{\alpha^{'}\gamma} )  \nn\\
&+p^{\alpha^{'}} (\eta^{\beta\gamma}\eta^{\alpha \gamma^{'}} +
\eta^{\beta\gamma^{'}} \eta^{\alpha \gamma} ) \nn\\
&-p^{\gamma} (\eta^{\alpha\beta}\eta^{\alpha^{'} \gamma^{'}} +
\eta^{\alpha^{'}\beta} \eta^{\alpha \gamma^{'}} ) \nn\\
&-p^{\gamma^{'}} (\eta^{\alpha\beta}\eta^{\alpha^{'} \gamma } +
\eta^{\alpha^{'}\beta} \eta^{\alpha \gamma} ) ]
\eeqa
in order to get symmetric expression:
\beqa
F^{\alpha \alpha^{'} \beta \gamma\gamma^{'}}_S (k  ,p ,q) =
-   {i\over 2} (2\alpha^{'})^3
[&+(q-k)^{\beta}  (\eta^{\alpha\gamma}\eta^{\alpha^{'}\gamma^{'}} +
\eta^{\alpha\gamma^{'}} \eta^{\alpha^{'}\gamma} ) \nn\\
&+(p- q)^{\alpha} (\eta^{\beta\gamma}\eta^{\alpha^{'}\gamma^{'}} +
\eta^{\beta\gamma^{'}} \eta^{\alpha^{'}\gamma} )  \nn\\
&+(p- q)^{\alpha^{'}} (\eta^{\beta\gamma}\eta^{\alpha \gamma^{'}} +
\eta^{\beta\gamma^{'}} \eta^{\alpha \gamma} ) \nn\\
&+(k-p)^{\gamma} (\eta^{\alpha\beta}\eta^{\alpha^{'} \gamma^{'}} +
\eta^{\alpha^{'}\beta} \eta^{\alpha \gamma^{'}} ) \nn\\
&+(k-p)^{\gamma^{'}} (\eta^{\alpha\beta}\eta^{\alpha^{'} \gamma } +
\eta^{\alpha^{'}\beta} \eta^{\alpha \gamma} ) ].
\eeqa
Substituting this into the expression (\ref{vecror2tensorvertex})
with the terms in the reversed cyclic orientation
$a,(\alpha,\alpha^{'}),k \leftrightarrow c,(\gamma,\gamma^{'}),q$
we shall get:
\beqa
\CV^{\alpha \alpha^{'} \beta \gamma\gamma^{'}}_{a  b c } (k  ,p ,q) =
 ~tr([\lambda^{a } , \lambda^{b}]\lambda^{c})
F^{\alpha \alpha^{'} \beta \gamma\gamma^{'}}_S (k  ,p ,q)~.
\eeqa
This expression should be compared with the expression (\ref{ymsymmetricvertex})
in tensor gauge field theory. We see that they have a similar Lorentz
structure, but with some differences in the coefficients. We don't know  exactly the
origin of this difference, but most probably it is connected with contributions
of higher rank non-Abelian tensor gauge fields, which we do not take into
consideration in this article.

\subsection{\it Tree-Level Amplitude for Two Anti-Symmetric Tensors and Vector}
The vertex operator for the antisymmetric $T_A$ rank-2 tensor boson on the forth level
is:
\be\label{antisymmetricvertex}
\zeta_{\alpha\alpha^{'}}(k):\ddot{X}^{\{\alpha}\dot{X}^{\alpha^{'}\}} e^{ik  X}
=\zeta_{\alpha\alpha^{'}}(k){1\over 2}:(\ddot{X}^{ \alpha}\dot{X}^{\alpha^{'} }-
\ddot{X}^{\alpha^{'} }\dot{X}^{\alpha })  e^{ik  X} :.
\ee
The antisymmetric wave functions of the pair of tensor gauge bosons
and the vector boson are:
\be
\zeta_{\alpha \alpha^{'}}(k),~~~\zeta_{\gamma\gamma^{'}}(q),~~~e_{ \beta}(p)
\ee
and we shall define for convenience
$k_1 \equiv k, k_2 \equiv q, k_3 \equiv p,$.
The mass-shell conditions are
$$\alpha^{'} k^{2}_{1}=\alpha^{'} k^{2}_{2}= -2,~   \alpha^{'} k^{2}_{3}=0$$
and  $k_1 +k_2 +k_3 =0$.
We have to calculate the correlation function:
\beqa
&F^{\alpha \alpha^{'} \beta \gamma\gamma^{'}}_A (k  ,p ,q) = \nn\\
\\
&=\int \prod_i ~d \mu(y_i)
<:\ddot{X}^{\{\alpha}\dot{X}^{\alpha^{'}\}} e^{ik  X}(y_1):
:\ddot{X}^{\{\gamma}\dot{X}^{\gamma^{'}\}} e^{iq X}(y_2):
:\dot{X}^{\beta}e^{ip X}(y_3):>  \nn\\
\\
& =\lim_{y_1=0,y_2=1,y_3 \rightarrow \infty}
\prod_{i<j}\vert y_i -y_j \vert^{2\alpha^{'}k_ik_j}~ y^{2}_{3}~ {1 \over 4}~~~~
\{ \CO(\alpha^{'})^5 (k)^5 + \CO(\alpha^{'})^4 (k)^3 +\nn\\
&+(2\alpha^{'})^2 F^{\beta}_{y_3} [ +
     {-6 \eta^{\alpha \gamma  }\over (y_1 -y_2)^4}
     {\eta^{\alpha^{'}\gamma^{'} }\over (y_1 -y_2)^2} +
     {-2 \eta^{\alpha \gamma^{'} }\over (y_1 -y_2)^3}
     {2 \eta^{\alpha^{'} \gamma }\over (y_1 -y_2)^3}  ~~~~~~~~~~~~~~~~  \nn\\
&-     {-2\eta^{\alpha \gamma  }\over (y_1 -y_2)^3}
     {2\eta^{\alpha^{'}\gamma^{'} }\over (y_1 -y_2)^3} -
     {-6 \eta^{\alpha \gamma^{'} }\over (y_1 -y_2)^4}
     { \eta^{\alpha^{'} \gamma }\over (y_1 -y_2)^2} \nn\\
&-     {2\eta^{\alpha \gamma  }\over (y_1 -y_2)^3}
     {-2\eta^{\alpha^{'}\gamma^{'} }\over (y_1 -y_2)^3} -
     { \eta^{\alpha \gamma^{'} }\over (y_1 -y_2)^2}
     { -6\eta^{\alpha^{'} \gamma }\over (y_1 -y_2)^4} \nn\\
&+     {\eta^{\alpha \gamma  }\over (y_1 -y_2)^2}
     {-6\eta^{\alpha^{'}\gamma^{'} }\over (y_1 -y_2)^4} +
     {2 \eta^{\alpha \gamma^{'} }\over (y_1 -y_2)^3}
     {-2 \eta^{\alpha^{'} \gamma }\over (y_1 -y_2)^3} ]\nn\\
&+  F^{\alpha}_{y_1}[-
     {2\eta^{\beta \gamma  }\over (y_3 -y_2)^3}
     {-2 \eta^{\alpha^{'}\gamma^{'} }\over (y_1 -y_2)^3} -
     {\eta^{ \beta\gamma^{'} }\over (y_3 -y_2)^2}
     {-6\eta^{\alpha^{'} \gamma }\over (y_1 -y_2)^4}  ~~~~~~~~~~~~~ \nn\\
&+ {\eta^{\beta \gamma  }\over (y_3 -y_2)^2}
     {-6 \eta^{\alpha^{'}\gamma^{'} }\over (y_1 -y_2)^4} +
     {2\eta^{ \beta\gamma^{'} }\over (y_3 -y_2)^3}
     {-2\eta^{\alpha^{'} \gamma }\over (y_1 -y_2)^3}]   \nn\\
&+  \dot{F}^{\alpha}_{y_1}[+
     {2\eta^{\beta \gamma  }\over (y_3 -y_2)^3}
     {\eta^{\alpha^{'}\gamma^{'} }\over (y_1 -y_2)^2} +
     {\eta^{ \beta\gamma^{'} }\over (y_3 -y_2)^2}
     {2\eta^{\alpha^{'} \gamma }\over (y_1 -y_2)^3}  ~~~~~~~~~~~~~ \nn\\
&- {\eta^{\beta \gamma  }\over (y_3 -y_2)^2}
     {2 \eta^{\alpha^{'}\gamma^{'} }\over (y_1 -y_2)^3} +
     {2\eta^{ \beta\gamma^{'} }\over (y_3 -y_2)^3}
     {\eta^{\alpha^{'} \gamma }\over (y_1 -y_2)^2} ]  \nn\\
&+  F^{\alpha^{'}}_{y_1}[+
     {2\eta^{\beta \gamma  }\over (y_3 -y_2)^3}
     {-2 \eta^{\alpha \gamma^{'} }\over (y_1 -y_2)^3} +
     {\eta^{ \beta\gamma^{'} }\over (y_3 -y_2)^2}
     {-6\eta^{\alpha  \gamma }\over (y_1 -y_2)^4}  ~~~~~~~~~~~~~ \nn\\
&- {\eta^{\beta \gamma  }\over (y_3 -y_2)^2}
     {-6 \eta^{\alpha \gamma^{'} }\over (y_1 -y_2)^4} -
     {2\eta^{ \beta\gamma^{'} }\over (y_3 -y_2)^3}
     {-2\eta^{\alpha  \gamma }\over (y_1 -y_2)^3}]   \nn\\
&+  \dot{F}^{\alpha^{'}}_{y_1}[-
     {2\eta^{\beta \gamma  }\over (y_3 -y_2)^3}
     {\eta^{\alpha \gamma^{'} }\over (y_1 -y_2)^2} -
     {\eta^{ \beta\gamma^{'} }\over (y_3 -y_2)^2}
     {2\eta^{\alpha  \gamma }\over (y_1 -y_2)^3}  ~~~~~~~~~~~~~ \nn\\
&+ {\eta^{\beta \gamma  }\over (y_3 -y_2)^2}
     {2 \eta^{\alpha \gamma^{'} }\over (y_1 -y_2)^3} +
     {2\eta^{ \beta\gamma^{'} }\over (y_3 -y_2)^3}
     {\eta^{\alpha  \gamma }\over (y_1 -y_2)^2} ]  \nn\\
&+  F^{\gamma }_{y_2}[~......................................
...........................................~]~~~~  \nn\\
&+ \dot{F}^{\gamma}_{y_2}[~...................................
...........................................~]~~~~ \nn\\
&+  F^{\gamma^{'} }_{y_2}[~......................................
...........................................~]~~~~  \nn\\
&+ \dot{F}^{\gamma^{'}}_{y_2}[~...................................
...........................................~]~~~~  \},
\eeqa
where dots denote the terms which one can get by interchanging
$\alpha, \alpha^{'}$ with $\gamma, \gamma^{'}$.
This amplitude has the same dimensional structure as the symmetric one:
\be
F^{\alpha \alpha^{'} \beta \gamma\gamma^{'}} \sim
 (\alpha^{'})^5 (k)^5 + (\alpha^{'})^4 (k)^3 + (\alpha^{'})^3 (k)^1
\ee
and we shall calculate the term which contains only the
first power of momentum, that is, the last term. Taking the corresponding limit
\beqa
& \lim_{y_1=0,~y_2=1,~y_3 \rightarrow \infty}
\prod_{i<j}\vert y_i -y_j \vert^{2\alpha^{'}k_ik_j}~\rightarrow 1\nn\\
&F^{\alpha}_{y_1} = -2 i \alpha^{'}~ ({p^{\alpha}  \over y_1-y_3}
+{q^{\alpha}  \over y_1-y_2})  \rightarrow -2 i \alpha^{'}~( - q^{ \alpha  } ) \nn\\
& F^{\gamma}_{y_2}= -2 i \alpha^{'}~({p^{\gamma}  \over y_2-y_3 }
+{k^{\gamma} \over y_2-y_1}) \rightarrow -2 i \alpha^{'}~( k^{\gamma  } ) \nn\\
&F^{\beta}_{y_3}= -2 i \alpha^{'}~({k^{\beta}  \over y_3-y_1}
+{q^{\beta}  \over y_3-y_2}) \rightarrow  -2 i \alpha^{'}~(-{k^{\beta}  \over y^2_3} )\nn\\
&\dot{F}^{\alpha}_{y_1} =  2 i \alpha^{'}~ ({p^{\alpha}  \over (y_1-y_3)^2}
+{q^{\alpha}  \over (y_1-y_2)^2})  \rightarrow  -2 i \alpha^{'}~(- q^{ \alpha  } ) \nn\\
& \dot{F}^{\gamma}_{y_2}=   2 i \alpha^{'}~({p^{\gamma}  \over (y_2-y_3)^2 }
+{k^{\gamma} \over (y_2-y_1)^2}) \rightarrow  - 2 i \alpha^{'}~( -k^{\gamma  } ) \nn\\
\eeqa
we shall get
\beqa
F^{\alpha \alpha^{'} \beta \gamma\gamma^{'}}_A (k  ,p ,q) =- i(2\alpha^{'})^3
[&+k^{\beta}  (\eta^{\alpha\gamma}\eta^{\alpha^{'}\gamma^{'}} -
\eta^{\alpha\gamma^{'}} \eta^{\alpha^{'}\gamma} ) \nn\\
&+q^{\alpha} (\eta^{\beta\gamma}\eta^{\alpha^{'}\gamma^{'}} -
\eta^{\beta\gamma^{'}} \eta^{\alpha^{'}\gamma} )  \nn\\
&- q^{\alpha^{'}} (\eta^{\beta\gamma}\eta^{\alpha \gamma^{'}} -
\eta^{\beta\gamma^{'}} \eta^{\alpha \gamma} ) \nn\\
&+k^{\gamma} (\eta^{\alpha\beta}\eta^{\alpha^{'} \gamma^{'}} -
\eta^{\alpha^{'}\beta} \eta^{\alpha \gamma^{'}} ) \nn\\
&-k^{\gamma^{'}} (\eta^{\alpha\beta}\eta^{\alpha^{'} \gamma } -
\eta^{\alpha^{'}\beta} \eta^{\alpha \gamma} ) ].
\eeqa
Adding the equal term
\beqa
F^{\alpha \alpha^{'} \beta \gamma\gamma^{'}}_A (k  ,p ,q) =- i(2\alpha^{'})^3
[&-q^{\beta}  (\eta^{\alpha\gamma}\eta^{\alpha^{'}\gamma^{'}} -
\eta^{\alpha\gamma^{'}} \eta^{\alpha^{'}\gamma} ) \nn\\
&-p^{\alpha} (\eta^{\beta\gamma}\eta^{\alpha^{'}\gamma^{'}} -
\eta^{\beta\gamma^{'}} \eta^{\alpha^{'}\gamma} )  \nn\\
&+p^{\alpha^{'}} (\eta^{\beta\gamma}\eta^{\alpha \gamma^{'}} -
\eta^{\beta\gamma^{'}} \eta^{\alpha \gamma} ) \nn\\
&-p^{\gamma} (\eta^{\alpha\beta}\eta^{\alpha^{'} \gamma^{'}} -
\eta^{\alpha^{'}\beta} \eta^{\alpha \gamma^{'}} ) \nn\\
&+p^{\gamma^{'}} (\eta^{\alpha\beta}\eta^{\alpha^{'} \gamma } -
\eta^{\alpha^{'}\beta} \eta^{\alpha \gamma} ) ]
\eeqa
we shall get a symmetric in momenta expression
\beqa
F^{\alpha \alpha^{'} \beta \gamma\gamma^{'}}_A (k  ,p ,q) =  - {i\over 2} (2\alpha^{'})^3
[&-(q-k)^{\beta}  (\eta^{\alpha\gamma}\eta^{\alpha^{'}\gamma^{'}} -
\eta^{\alpha\gamma^{'}} \eta^{\alpha^{'}\gamma} ) \nn\\
&-(p- q)^{\alpha} (\eta^{\beta\gamma}\eta^{\alpha^{'}\gamma^{'}} -
\eta^{\beta\gamma^{'}} \eta^{\alpha^{'}\gamma} )  \nn\\
&+(p- q)^{\alpha^{'}} (\eta^{\beta\gamma}\eta^{\alpha \gamma^{'}} -
\eta^{\beta\gamma^{'}} \eta^{\alpha \gamma} ) \nn\\
&-(p-k)^{\gamma} (\eta^{\alpha\beta}\eta^{\alpha^{'} \gamma^{'}} -
\eta^{\alpha^{'}\beta} \eta^{\alpha \gamma^{'}} ) \nn\\
&+(p-k)^{\gamma^{'}} (\eta^{\alpha\beta}\eta^{\alpha^{'} \gamma } -
\eta^{\alpha^{'}\beta} \eta^{\alpha \gamma} ) ].
\eeqa
Substituting this into the expression (\ref{vecror2tensorvertex})
with the terms in the reversed cyclic orientation
$a,(\alpha,\alpha^{'}),k \leftrightarrow c,(\gamma,\gamma^{'}),q$
we shall finally get:
\beqa
\CV^{\alpha \alpha^{'} \beta \gamma\gamma^{'}}_{a  b c } (k  ,p ,q) =
tr([\lambda^{a } , \lambda^{b}]\lambda^{c})
F^{\alpha \alpha^{'} \beta \gamma\gamma^{'}}_A (k  ,p ,q).
\eeqa
This expression should be compared with the expression (\ref{ymantisymmetricvertex}) and
again we see that they have a similar structure.

\subsection{\it Mixed Symmetry Amplitudes}

Finally we shall calculate the mixed amplitude between symmetric and antisymmetric
tensor bosons and a vector. The vertex operator for the symmetric $T_S$ and antisymmetric $T_A$
rank-2 tensor bosons have been given above (\ref{symmetricvertex}) and (\ref{antisymmetricvertex}).
Therefore  the matrix element is:
\beqa
&F^{\alpha \alpha^{'} \beta \gamma\gamma^{'}}_{SA} (k  ,p ,q) = \nn\\
&=\int \prod_i ~d \mu(y_i)
<:\dot{X}^{\alpha}\dot{X}^{\alpha^{'}} e^{ik  X}(y_1):
:\ddot{X}^{\{\gamma}\dot{X}^{\gamma^{'}\}} e^{iq X}(y_2):
:\dot{X}^{\beta}e^{ip X}(y_3):>  \nn\\
& =\lim_{y_1=0,y_2=1,y_3 \rightarrow \infty}
\prod_{i<j}\vert y_i -y_j \vert^{2\alpha^{'}k_ik_j}~ y^{2}_{3}~~ {1 \over 2}~~~
\{ \CO(\alpha^{'})^5 (k)^5 + \CO(\alpha^{'})^4 (k)^3 +\nn\\
&+(2\alpha^{'})^2 F^{\beta}_{y_3} [ +
     {2 \eta^{\alpha \gamma  }\over (y_1 -y_2)^3}
     {\eta^{\alpha^{'}\gamma^{'} }\over (y_1 -y_2)^2} +
     {  \eta^{\alpha \gamma^{'} }\over (y_1 -y_2)^2}
     {2 \eta^{\alpha^{'} \gamma }\over (y_1 -y_2)^3}  ~~~~~~~~~~~~~~~~  \nn\\
&-     { \eta^{\alpha \gamma  }\over (y_1 -y_2)^2}
     {2\eta^{\alpha^{'}\gamma^{'} }\over (y_1 -y_2)^3} -
     {2 \eta^{\alpha \gamma^{'} }\over (y_1 -y_2)^3}
     { \eta^{\alpha^{'} \gamma }\over (y_1 -y_2)^2}] \nn\\
&+  F^{\alpha}_{y_1}[+
     {-2\eta^{\beta \gamma  }\over (y_2 -y_3)^3}
     {\eta^{\alpha^{'}\gamma^{'} }\over (y_1 -y_2)^2} +
     {\eta^{ \beta\gamma^{'} }\over (y_2 -y_3)^2}
     {2\eta^{\alpha^{'} \gamma }\over (y_1 -y_2)^3}  ~~~~~~~~~~~~~ \nn\\
&- {\eta^{\beta \gamma  }\over (y_2 -y_3)^2}
     {2\eta^{\alpha^{'}\gamma^{'} }\over (y_1 -y_2)^3} -
     {-2\eta^{ \beta\gamma^{'} }\over (y_2 -y_3)^3}
     {\eta^{\alpha^{'} \gamma }\over (y_1 -y_2)^2}]   \nn\\
&+  F^{\alpha^{'}}_{y_1}[+
     {-2\eta^{\beta \gamma  }\over (y_2 -y_3)^3}
     { \eta^{\alpha \gamma^{'} }\over (y_1 -y_2)^2} +
     {\eta^{ \beta\gamma^{'} }\over (y_2 -y_3)^2}
     {2\eta^{\alpha  \gamma }\over (y_1 -y_2)^3}  ~~~~~~~~~~~~~ \nn\\
&- {\eta^{\beta \gamma  }\over (y_2 -y_3)^2}
     {2 \eta^{\alpha \gamma^{'} }\over (y_1 -y_2)^3} -
     {-2\eta^{ \beta\gamma^{'} }\over (y_2 -y_3)^3}
     {\eta^{\alpha  \gamma }\over (y_1 -y_2)^2}]   \nn\\
&+  \dot{F}^{\gamma}_{y_2}[+
     {\eta^{\alpha \gamma^{'}  }\over (y_1 -y_2)^2}
     {\eta^{\alpha^{'} \beta }\over (y_1 -y_3)^2} +
     {\eta^{ \alpha \beta}\over (y_1 -y_3)^2}
     {\eta^{\alpha^{'}  \gamma^{'} }\over (y_1 -y_2)^2}]  ~~~~~~~~~~~~~ \nn\\
&+ F^{\gamma}_{y_2}[ - {2\eta^{\alpha \gamma^{'}  }\over (y_1 -y_2)^3}
     {\eta^{\alpha^{'} \beta }\over (y_1 -y_3)^2} -
     {\eta^{\alpha \beta  }\over (y_1 -y_3)^2}
     {2\eta^{\alpha^{'}  \gamma^{'} }\over (y_1 -y_2)^3} ] ~~~~~~~~~~~~~ \nn\\
&+  F^{\gamma^{'}}_{y_2}[+
     {2\eta^{\alpha \gamma   }\over (y_1 -y_2)^3}
     {\eta^{\alpha^{'} \beta }\over (y_1 -y_3)^2} +
     {\eta^{ \alpha\beta }\over (y_1 -y_3)^2}
     {2\eta^{\alpha^{'}  \gamma }\over (y_1 -y_2)^3}]  ~~~~~~~~~~~~~ \nn\\
&+ \dot{F}^{\gamma^{'}}_{y_2}[ - {\eta^{\alpha \gamma }\over (y_1 -y_2)^2}
     {\eta^{\alpha^{'} \beta }\over (y_1 -y_3)^2} -
     {\eta^{\alpha \beta  }\over (y_1 -y_3)^2}
     {\eta^{\alpha^{'}  \gamma}\over (y_1 -y_2)^2} ] ~  \}.
\eeqa
Taking the corresponding limit we shall get
\beqa
F^{\alpha \alpha^{'} \beta \gamma\gamma^{'}}_{S~~~~A} (k  ,p ,q) =- i(2\alpha^{'})^3
[&-q^{\alpha} (\eta^{\beta\gamma}\eta^{\alpha^{'}\gamma^{'}} -
\eta^{\beta\gamma^{'}} \eta^{\alpha^{'}\gamma} )  \nn\\
&- q^{\alpha^{'}} (\eta^{\beta\gamma}\eta^{\alpha \gamma^{'}} -
\eta^{\alpha \gamma}\eta^{\beta\gamma^{'}}  ) \nn\\
&+{1\over 2}k^{\gamma} (\eta^{\alpha\beta}\eta^{\alpha^{'} \gamma^{'}} +
\eta^{\alpha \gamma^{'}}\eta^{\alpha^{'}\beta}  ) \nn\\
&-{1\over 2}k^{\gamma^{'}} (\eta^{\alpha\beta}\eta^{\alpha^{'} \gamma } +
\eta^{\alpha \gamma}\eta^{\alpha^{'}\beta}  ) ]
\eeqa
or, in equivalent form, as
\beqa
F^{\alpha \alpha^{'} \beta \gamma\gamma^{'}}_{S~~~~A} (k  ,p ,q) =- {i\over 2}(2\alpha^{'})^3
[&+(p-q)^{\alpha} (\eta^{\beta\gamma}\eta^{\alpha^{'}\gamma^{'}} -
\eta^{\beta\gamma^{'}} \eta^{\alpha^{'}\gamma} )  \nn\\
&+(p- q)^{\alpha^{'}} (\eta^{\beta\gamma}\eta^{\alpha \gamma^{'}} -
\eta^{\alpha \gamma}\eta^{\beta\gamma^{'}}  ) \nn\\
&+{1\over 2}(k-p)^{\gamma} (\eta^{\alpha\beta}\eta^{\alpha^{'} \gamma^{'}} +
\eta^{\alpha \gamma^{'}}\eta^{\alpha^{'}\beta}  ) \nn\\
&-{1\over 2}(k-p)^{\gamma^{'}} (\eta^{\alpha\beta}\eta^{\alpha^{'} \gamma } +
\eta^{\alpha \gamma}\eta^{\alpha^{'}\beta}  ) ].
\eeqa
The mixed symmetry vertex will take the following form:
\beqa
F^{\alpha \alpha^{'} \beta \gamma\gamma^{'}}_{A~~~~S} (k  ,p ,q) =
- {i\over 2}(2\alpha^{'})^3
[&+{1\over 2}(p-q)^{\alpha} (\eta^{\beta\gamma}\eta^{\alpha^{'}\gamma^{'}} +
\eta^{\beta\gamma^{'}} \eta^{\alpha^{'}\gamma} )  \nn\\
&-{1\over 2}(p- q)^{\alpha^{'}} (\eta^{\beta\gamma}\eta^{\alpha \gamma^{'}} +
\eta^{\alpha \gamma}\eta^{\beta\gamma^{'}}  ) \nn\\
&+(k-p)^{\gamma} (\eta^{\alpha\beta}\eta^{\alpha^{'} \gamma^{'}} -
\eta^{\alpha \gamma^{'}}\eta^{\alpha^{'}\beta}  ) \nn\\
&+(k-p)^{\gamma^{'}} (\eta^{\alpha\beta}\eta^{\alpha^{'} \gamma } -
\eta^{\alpha \gamma}\eta^{\alpha^{'}\beta}  ) ]
\eeqa
Plugging  both expressions into (\ref{vecror2tensorvertex})
with the terms in the reversed cyclic orientation
$a,(\alpha,\alpha^{'}),k \leftrightarrow c,(\gamma,\gamma^{'}),q$
we shall get:
\beqa
\CV^{\alpha \alpha^{'} \beta \gamma\gamma^{'}}_{a  b c } (k  ,p ,q) =
 ~tr([\lambda^{a } , \lambda^{b}]\lambda^{c})~[~
F^{\alpha \alpha^{'} \beta \gamma\gamma^{'}}_{S~~~~A} (k  ,p ,q)+
F^{\alpha \alpha^{'} \beta \gamma\gamma^{'}}_{A~~~~S} (k  ,p ,q)~]
\eeqa
which is in agreement with the expression we got in the tensor gauge field
theory (\ref{mixedsymmetryvertex1}) and (\ref{mixedsymmetryvertex2}).

\section{{\it Scattering Amplitude of Two Scalars and Two Tensors }}

We shall get less trivial information about the structure of the VTT vertex of
the open string theory if we shall compute the four particle scattering amplitude
of two tachyon  and two tensor bosons (see Fig.\ref{fig4}). In the low energy limit these amplitudes
will be dominated by the exchange of the massless vector boson (see Fig.\ref{fig5}) and shall
provide the information about the structure of cubic vertices $\CV_{1-2-2}$ and $\CV_{0-1-2}$.
We are interested to calculate the following scattering amplitude on the disk (see Fig.\ref{fig4}):
\beqa
&F^{\mu\nu,\lambda\rho}  (k_1,k_2;k_3,k_4) = \nn\\
&=\int \prod_i ~d \mu(y_i)
<:e^{ik_1  X}(y_1):
:e^{ik_2 X}(y_2):
:\dot{X}^{\mu}\dot{X}^{\nu } e^{ik_3 X}(y_3):
:\dot{X}^{\lambda}\dot{X}^{\rho } e^{ik_4 X}(y_4):>
                            \nn\\
& =\int ~d \mu(y_i)
\prod_{i<j}\vert y_i -y_j \vert^{2\alpha^{'}k_ik_j} ~\nn\\
&\{~ (F^{ \mu}F^{ \nu})_{y_3}
(F^{ \lambda}F^{\rho  })_{y_4}                  +        \nn\\
&+ (- 2\alpha^{'})
[ F^{\mu}_{y_3} F^{\lambda}_{y_4}
      {\eta^{\nu\rho }\over (y_3 -y_4)^2} +
F^{\mu}_{y_3} F^{\rho}_{y_4}
         {\eta^{\nu\lambda}\over (y_3 -y_4)^2} +
F^{\nu }_{y_3} F^{\lambda}_{y_4}
         {\eta^{\mu\rho }\over (y_3 -y_4)^2} +
F^{\nu }_{y_3} F^{\rho}_{y_4}
       {\eta^{\mu\lambda}\over (y_3 -y_4)^2}] + \nn\\
&+ (-2\alpha^{'})^2[
     {\eta^{\mu\lambda}\over (y_3 -y_4)^2}
     {\eta^{\nu\rho }\over (y_3 -y_4)^2} +
     {\eta^{ \mu\rho }\over (y_3 -y_4)^2}
     {\eta^{\nu\lambda }\over (y_3 -y_4)^2} ]   \}.
\eeqa
We shall fix the integration measure by the choice
$ y_4=0,y_2=1,y_1 \rightarrow \infty  $.
The mass-shell conditions are
$$\alpha^{'} k^{2}_{1}=\alpha^{'} k^{2}_{2}= +1,~~~~~
\alpha^{'} k^{2}_{3}=\alpha^{'} k^{2}_{4}=-1$$
and  $k_1 +k_2 +k_3 +k_4 =0$. Thus
\beqa\label{intgral}
&\int ~d \mu(y_i)
\prod_{i<j}\vert y_i -y_j \vert^{2\alpha^{'}k_ik_j}~=
\int^{+\infty}_{-\infty} dy_3~y^{2}_{1}
\prod^{ i<j}_{y_4=0,~y_2=1,~y_1 \rightarrow \infty }
\vert y_i -y_j \vert^{2\alpha^{'}k_ik_j}~\rightarrow \nn\\
&=
\int^{+\infty}_{-\infty}~ dy_3~
\vert y_3\vert^{2\alpha^{'}k_3k_4}~\vert 1- y_3\vert^{2\alpha^{'}k_2k_3}
\eeqa
and
\beqa
&F^{\mu }_{y_3} = -2 i \alpha^{'}~ ({k^{\mu }_1 \over y_3-y_1}
+{k^{\mu }_2 \over y_3-y_2}+{k^{\mu }_4 \over y_3-y_4})  \rightarrow
-2 i \alpha^{'}~ ({k^{\mu }_2 \over y_3-1}+{k^{\mu }_4 \over y_3}  ),
\nn\\
&F^{\mu }_{y_4} = -2 i \alpha^{'}~ ({k^{\mu }_1 \over y_4-y_1}
+{k^{\mu }_2 \over y_4-y_2}+{k^{\mu }_3 \over y_4-y_3})  \rightarrow
-2 i \alpha^{'}~ ( -  k^{\mu }_2 -{ k^{\mu }_3 \over y_3  }   ).
\eeqa
\begin{figure}
\centerline{\hbox{\psfig{figure=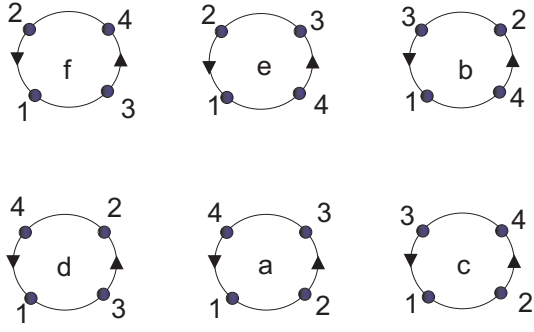,width=7cm}}}
\caption[fig4]{Orderings of four open string vertex operators
on the disk. The integral over $y_3$ in (\ref{intgral}) splits into
three ranges $[-\infty, 0],[0,1],[1,+\infty]$.
For these three ranges the vertex operators are ordered as in figures
(d),(a) and (c) respectively. For the reversed
cyclic permutation we shall get figures (f) (e) and (b).
The coordinate $y_3$ increases in the direction of arrow.
}
\label{fig4}
\end{figure}
Therefore we shall get
\beqa\label{matrixelement}
&F^{\mu\nu,\lambda\rho}  (k_1,k_2,k_3,k_4) =
 \int^{+\infty}_{-\infty}~ dy_3~
\vert y_3\vert^{2\alpha^{'}k_3k_4}~\vert 1- y_3\vert^{2\alpha^{'}k_2k_3} \nn\\
&  \{~~ (2\alpha^{'})^4 (  { k^{\mu }_2 \over y_3-1}+{k^{\mu }_4 \over y_3}  )
( { k^{\nu }_2 \over y_3-1  }+  {k^{\nu }_4 \over y_3}  )
(   k^{\lambda }_2 +{ k^{\lambda}_3 \over y_3  }   )
(   k^{\rho }_2 +{ k^{\rho }_3 \over y_3  }   )  \nn\\
&-(2\alpha^{'})^3~{1\over y^{2}_{3}}~[
(  { k^{\mu }_2 \over y_3-1}+{k^{\mu }_4 \over y_3}  )
((   k^{\rho }_2 +{ k^{\rho }_3 \over y_3  } ) \eta^{\nu\lambda}
+  (k^{\lambda }_2 +{ k^{\lambda}_3 \over y_3  } )\eta^{\nu\rho} ))\nn\\
&+(  { k^{\nu}_2 \over y_3-1}+{k^{\nu }_4 \over y_3}  )
((   k^{\lambda }_2 +{ k^{\lambda }_3 \over y_3  } ) \eta^{\mu\rho}
+  (k^{\rho }_2 +{ k^{\rho}_3 \over y_3  } ) \eta^{\mu\lambda} )  ]\nn\\
&+(2\alpha^{'})^2 ~{1\over y^{4}_{3}}~[\eta^{\mu\lambda}\eta^{\nu\rho}
+\eta^{\mu\rho}\eta^{\nu\lambda}]~~
\}
\eeqa
The integration over $y_3$ can be divided into three regions
$[-\infty, 0],[0,1],[1,+\infty]$. These pieces can be depicted
by three diagrams d) a) and c) on the Fig. \ref{fig4}. For the reversed
cyclic permutation we shall get diagrams f) e) and b) on the Fig. \ref{fig4}.
Introducing Mandelshtam variables
$$
s=-(k_1 +k_2)^2,~~~t=-(k_2 +k_3)^2,~~~u=-(k_2 +k_4)^2
$$
we can represent the contribution of the (s,t) diagrams a) and e) in the form
\beqa
&-(2\alpha^{'})^3~(tr(\lambda^{a_1}\lambda^{a_2}\lambda^{a_3}\lambda^{a_4})+
tr(\lambda^{a_4}\lambda^{a_3}\lambda^{a_2}\lambda^{a_1}))\nn\\
&[B(-\alpha^{'}s ,-\alpha^{'}t+1 ) K^{\mu\rho\nu\lambda}(k_4,k_2)+
B(-\alpha^{'}s -1,-\alpha^{'}t +1 ) K^{\mu\rho\nu\lambda}(k_4,k_3)\nn\\
&-B(-\alpha^{'}s +1,-\alpha^{'}t ) K^{\mu\rho\nu\lambda}(k_2,k_2)-
B(-\alpha^{'}s ,-\alpha^{'}t ) K^{\mu\rho\nu\lambda}(k_2,k_3)],
\eeqa
the contribution of the (s,u) diagrams f) and c) in the form
\beqa
&-(2\alpha^{'})^3~(tr(\lambda^{a_1}\lambda^{a_2}\lambda^{a_4}\lambda^{a_3})+
tr(\lambda^{a_1}\lambda^{a_3}\lambda^{a_4}\lambda^{a_2}))\nn\\
&[-B(-\alpha^{'}s ,-\alpha^{'}u ) K^{\mu\rho\nu\lambda}(k_4,k_2)
+B(-\alpha^{'}s -1,-\alpha^{'}u +1 ) K^{\mu\rho\nu\lambda}(k_4,k_3)\nn\\
&-B(-\alpha^{'}s +1,-\alpha^{'}u ) K^{\mu\rho\nu\lambda}(k_2,k_2)
+B(-\alpha^{'}s ,-\alpha^{'}u+1 ) K^{\mu\rho\nu\lambda}(k_2,k_3)]
\eeqa
and the the contribution of the (u,t) diagrams b) and d) in the form
\beqa
&-(2\alpha^{'})^3~(tr(\lambda^{a_1}\lambda^{a_3}\lambda^{a_2}\lambda^{a_4})+
tr(\lambda^{a_1}\lambda^{a_4}\lambda^{a_2}\lambda^{a_3}))\nn\\
&[B(-\alpha^{'}u ,-\alpha^{'}t+1 ) K^{\mu\rho\nu\lambda}(k_4,k_2)
+B(-\alpha^{'}u+1,-\alpha^{'}t +1 ) K^{\mu\rho\nu\lambda}(k_4,k_3)\nn\\
&+B(-\alpha^{'}u,-\alpha^{'}t ) K^{\mu\rho\nu\lambda}(k_2,k_2)
+B(-\alpha^{'}u+1 ,-\alpha^{'}t ) K^{\mu\rho\nu\lambda}(k_2,k_3)],
\eeqa
where
\be
K^{\mu\rho\nu\lambda}(k ,p)= k^{\mu}(p^{\rho}\eta^{\nu\lambda} +
p^{\lambda}\eta^{\nu\rho}) + k^{\nu}(p^{\lambda}\eta^{\mu\rho} +
p^{\rho}\eta^{\mu\lambda}).
\ee
Considering the limit
$
\alpha^{'} s , \alpha^{'}t, \alpha^{'}u \rightarrow 0
$
of the Euler functions
we shall get for the s , t and u channel contributions:
\beqa
-(2\alpha^{'})^3~\{&+{1\over \alpha^{'}s }
tr[\lambda^{a_1},\lambda^{a_2}][\lambda^{a_3}\lambda^{a_4}]
(K(k_2,k_3) -K(k_4,k_2))^{\mu\rho\nu\lambda}\nn\\
&-{1\over \alpha^{'}t }
tr[\lambda^{a_1},\lambda^{a_4}][\lambda^{a_2}\lambda^{a_3}]
(K(k_2,k_2) + K(k_2,k_3))^{\mu\rho\nu\lambda} \nn\\
&-{1\over \alpha^{'}u }
tr[\lambda^{a_1},\lambda^{a_3}][\lambda^{a_2}\lambda^{a_4}]
(K(k_2,k_2) - K(k_4,k_2))^{\mu\rho\nu\lambda} \}.
\eeqa
which are shown on the Fig.\ref{fig5}.

\begin{figure}
\centerline{\hbox{\psfig{figure=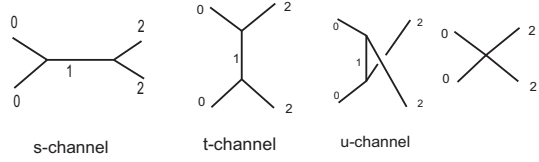,width=7cm}}}
\caption[fig5]{Feynman diagrams contributing to the low-energy limit
of the open string scattering amplitude.
The poles correspond to the exchange by a massless vector boson
in s,t and u channels. The scalars are depicted as 0, vectors as 1 and tensors as 2.
The quantities of main interest here are the dimensionless vertices $\CV_{1-2-2}$
between a vector and two tensors in s-channel and $\CV_{0-1-2}$ between a
scalar, vector and a tensor in t,u channels. The last graph represents the contact vertex
$\CV_{0-0-2-2}$.
}
\label{fig5}
\end{figure}
The last term in the matrix element (\ref{matrixelement}) has no momentum dependence
and represents the contact term. Evaluating the integration over $y_3$ in same
way as we did above one can get:
\beqa
 (2\alpha^{'})^2~\{&+(~tr \lambda^{a_1}\lambda^{a_2}\lambda^{a_3}\lambda^{a_4} +
tr \lambda^{a_4}\lambda^{a_3}\lambda^{a_2}\lambda^{a_1} ~)
B(-\alpha^{'}s-1 ,-\alpha^{'}t+1 )
\nn\\
&+(~tr\lambda^{a_1}\lambda^{a_2}\lambda^{a_4}\lambda^{a_3} +
tr\lambda^{a_1}\lambda^{a_3}\lambda^{a_4}\lambda^{a_2}~)
B(-\alpha^{'}s -1,-\alpha^{'}u +1 )
\nn\\
&+(~tr\lambda^{a_1}\lambda^{a_3}\lambda^{a_2}\lambda^{a_4})+
tr\lambda^{a_1}\lambda^{a_4}\lambda^{a_2}\lambda^{a_3}~)
B(-\alpha^{'}t +1,-\alpha^{'}u+1 ) \}
\nn\\
~&( \eta^{\mu\lambda} \eta^{\nu\rho} + \eta^{\mu\rho} \eta^{\nu \lambda} ).
\eeqa
Again considering the limit
$
\alpha^{'} s , \alpha^{'}t, \alpha^{'}u \rightarrow 0
$
of the Euler functions
we shall get for the contact term
\beqa
 (2\alpha^{'})^2~ \{~tr[\lambda^{a_1},\lambda^{a_4}][\lambda^{a_2},\lambda^{a_3}] +
{t\over s} ~tr[\lambda^{a_1},\lambda^{a_2}][\lambda^{a_3},\lambda^{a_4}] ~\}
 (\eta^{\mu\lambda} \eta^{\nu\rho} + \eta^{\mu\rho} \eta^{\nu \lambda} ).
\eeqa
This is the quartic vertex of $\CV_{0-0-2-2}$ and should be compared with
correspond vertex in non-Abelian tensor gauge field theory
\cite{Savvidy:2005fi,Savvidy:2005zm,Savvidy:2005ki}.

\section{{\it Conclusion}}

Our intention in this article was to compare the structure of the tree level
scattering amplitudes in non-Abelian tensor gauge field theory and in open string
theory with Chan-Paton charges. We limit ourselves considering only lower rank
tensor fields in both theories. We identify the symmetric $T_S$ and antisymmetric $T_A$ components of the
second rank tensor gauge field $A^{a}_{\alpha \acute{\alpha}}$ with the string excitations
$T_S$ and $T_A$ on the third and forth levels. In the process of this identification we have been
selecting only those parts of the tree level scattering amplitudes in the open string theory
which have {\it dimensionless coupling constants in four dimensions}. It should be mentioned
that not all amplitudes have these dimensionless parts. In particular, the scattering amplitude
of three tensor bosons $T$ does not have dimensionless parts. It seems that {\it dimensionless subclass
of tree level scattering amplitudes may provide information
about the structure of the open string theory at the deepest level}.

In this respect one could mention that
the cubic and quartic vertices of the massless Yang-Mills theory and a massive theory with spontaneous
symmetry breaking are identical. Therefore if we want to extract a genuine massless proto-theory
from the open string theory, which in accordance to  David Gross is in a broken phase \cite{Gross:1988ue},
it seems very natural to consider a subclass of amplitudes with dimensionless coupling
constant and compare them with the amplitudes in non-Abelian tensor gauge field theory.

It is also interesting to mention that the ratio of the masses of the tensor gauge bosons
$T_S$ and $T_A$ in non-Abelian tensor gauge theory with spontaneous
symmetry breaking is \cite{Savvidy:2008zy}
\be
{m^{2}_{A} \over m^{2}_{S} } = 3,
\ee
while in the open string theory it is (see Fig.\ref{fig1})
\be
{m^{2}_A \over m^2_S} = 4.
\ee
In both theories the antisymmetric tensor is more heavy.

In conclusion I would like to thank Costas Bachas for stimulating discussions and his suggestion
of comparing scattering amplitudes in both theories. I also would like to thank
Sebastian Guttenberg and Spiros Konitopoulos for many discussions and CERN Theory Division
for hospitality, where part of this work was completed.
This work was partially supported by ENRAGE (European Network on Random
Geometry), Marie Curie Research Training Network, contract MRTN-CT-2004-
005616.


\begin{thebibliography}{99}


\bibitem{Neveu:1971mu}
  A.~Neveu and J.~Scherk,
\emph{ Connection between Yang-Mills fields and dual models,}
  Nucl.\ Phys.\  B {\bf 36} (1972) 155.

\bibitem{Green:1987sp}
  M.~B.~Green, J.~H.~Schwarz and E.~Witten,
\emph{ Superstring theory. Vol. 1: Introduction,}
{\it  Cambridge, UK: Univ. Pr. ( 1987) 469 P. ( Cambridge Monographs On Mathematical Physics)}

\bibitem{Polchinski:1998rq}
  J.~Polchinski,
\emph{String theory. Vol. 1: An introduction to the bosonic string,}
{\it  Cambridge, UK: Univ. Pr. (1998) 402 p}


\bibitem{Gross:1988ue}
  D.~J.~Gross,
\emph{ High-Energy Symmetries Of String Theory},
  Phys.\ Rev.\ Lett.\  {\bf 60} (1988) 1229.

\bibitem{Gross:1987kz}
D.~J.~Gross and P.~F.~Mende,
\emph{ The High-Energy Behavior Of String Scattering Amplitudes},
Phys.\ Lett.\ B {\bf 197} (1987) 129.

\bibitem{Gross:1987ar}
D.~J.~Gross and P.~F.~Mende,
\emph{ String Theory Beyond The Planck Scale},
Nucl.\ Phys.\ B {\bf 303} (1988) 407.


\bibitem{Witten:1988zd}
E.~Witten,
\emph{ The Search For Higher Symmetry In String Theory},
Philos. Trans. R. Soc. London A320 (1989) 349-357;\\
E.~Witten,
\emph{ Space-Time And Topological Orbifolds},
Phys.\ Rev.\ Lett.\  {\bf 61} (1988) 670

\bibitem{Mende:1992pm}
P.~F.~Mende,
\emph{ String theory at short distance and the principle of equivalence},
arXiv:hep-th/9210001.

\bibitem{Mende:1989wt}
P.~F.~Mende and H.~Ooguri,
\emph{ Borel Summation Of String Theory For Planck Scale Scattering},
Nucl.\ Phys.\ B {\bf 339} (1990) 641.

\bibitem{Mende:1994wf}
P.~F.~Mende,
\emph{ High-energy string collisions in a compact space},
Phys.\ Lett.\ B {\bf 326} (1994) 216

\bibitem{Moore:1993ns}
G.~W.~Moore,
\emph{ Symmetries of the bosonic string S matrix},
arXiv:hep-th/9310026.


\bibitem{Savvidy:2003fx}
G.~K.~Savvidy,~
``Tensionless strings: Physical Fock space and higher spin fields,''
Int.\ J.\ Mod.\ Phys.\ A {\bf 19},  (2004) 3171-3194.


\bibitem{Giddings:2007bw}
  S.~B.~Giddings, D.~J.~Gross and A.~Maharana,
\emph{ Gravitational effects in ultrahigh-energy string scattering,}
 arXiv:hep-th/0705.1816


\bibitem{Savvidy:2005fe}
G.~Savvidy,
\emph{Tensionless strings, correspondence with SO(D,D) sigma model},
Phys.\ Lett.\ B {\bf 615} (2005) 285.



\bibitem{Amati:1988tn}
D.~Amati, M.~Ciafaloni and G.~Veneziano,
\emph{Can Space-Time Be Probed Below The String Size?}
Phys.\ Lett.\ B {\bf 216} (1989) 41.


\bibitem{Lindstrom:1990qb}
U.~Lindstrom, B.~Sundborg and G.~Theodoridis,
\emph{The Zero Tension Limit Of The Superstring,}
Phys.\ Lett.\ B {\bf 253} (1991) 319.

\bibitem{DeVega:1992tm}
H.~J.~De Vega and A.~Nicolaidis,
\emph{Strings in strong gravitational fields,}
Phys.\ Lett.\ B {\bf 295} (1992) 214.



\bibitem{Bakas:2004jq}
I.~Bakas and C.~Sourdis,
\emph{On the tensionless limit of gauged WZW models,}
arXiv:hep-th/0403165.



\bibitem{Witten:1985cc}
  E.~Witten,
  \emph{Noncommutative Geometry And String Field Theory,}
  Nucl.\ Phys.\  B {\bf 268} (1986) 253.


\bibitem{Thorn:1985fa}
  C.~B.~Thorn,
   \emph{Comments On Covariant Formulations Of String Theories,}
  Phys.\ Lett.\  {\bf 159B} (1985) 107;
    \emph{ String Field Theory,} Phys.\ Rep.\  {\bf 174C} (1989) 1


\bibitem{Siegel:1985tw}
W.~Siegel and B.~Zwiebach,
\emph{Gauge String Fields,}
Nucl.\ Phys.\ B {\bf 263} (1986) 105.




\bibitem{Arefeva:1989cp}
  I.~Y.~Arefeva, P.~B.~Medvedev and A.~P.~Zubarev,
  \emph{ New Representation For String Field Solves
  The Consistence Problem For Open Superstring Field,}
  Nucl.\ Phys.\  B {\bf 341} (1990) 464.



\bibitem{Siegel:1988yz}
  W.~Siegel,
 \emph{Introduction to string field theory,}
  arXiv:hep-th/0107094.

\bibitem{Taylor:2003gn}
  W.~Taylor and B.~Zwiebach,
   \emph{D-branes, tachyons, and string field theory,}
  arXiv:hep-th/0311017 (see formulas (201) and (202) of section 6.5).

\bibitem{Taylor:2006ye}
  W.~Taylor,
  \emph{String field theory,}
  arXiv:hep-th/0605202.








\bibitem{yang} C.N.Yang and R.L.Mills. "Conservation of Isotopic
Spin and Isotopic Gauge Invariance". Phys.\ Rev.\ {\bf 96} (1954) 191

\bibitem{Savvidy:2005fi}
G.~Savvidy,
\emph{Non-Abelian tensor gauge fields: Generalization of Yang-Mills theory},
Phys.\ Lett.\ B {\bf 625} (2005) 341


\bibitem{Savvidy:2005zm}
  G.~Savvidy,
 \emph{Non-abelian tensor gauge fields. I,}
  Int.\ J.\ Mod.\ Phys.\ A {\bf 21} (2006) 4931;


\bibitem{Savvidy:2005ki}
  G.~Savvidy,
  \emph{Non-abelian tensor gauge fields. II,}
  Int.\ J.\ Mod.\ Phys.\ A {\bf 21} (2006) 4959;

\bibitem{Savvidy:2005at}
G.~Savvidy and T.~Tsukioka,
\emph{Gauge invariant Lagrangian for non-Abelian tensor gauge fields of fourth
rank,} Prog.\ Theor.\ Phys.\ {\bf 117} (2007) 4;
arXiv:hep-th/0512344.

\bibitem{Barrett:2007}
J.~Barrett and G.~Savvidy
\emph{A Dual Lagrangian for Non-Abelian Tensor Gauge Fields},
Phys.\ Lett.\ B {\bf 652} (2007) 141-145

\bibitem{Guttenberg:2008zn}
  S.~Guttenberg and G.~Savvidy,
\emph{Duality transformation of non-Abelian tensor gauge fields,}
  Mod.\ Phys.\ Lett.\  A {\bf 23} (2008) 999

\bibitem{Konitopoulos:2008vv}
  S.~Konitopoulos, R.~Fazio and G.~Savvidy,
  \emph{Tensor gauge boson production in high energy collisions,}
  arXiv:0803.0075 [hep-th].


\bibitem{Konitopoulos:2008bd}
  S.~Konitopoulos and G.~Savvidy,
  \emph{Production of Spin-Two Gauge Bosons,}
  arXiv:0804.0847 [hep-th].



\bibitem{Paton:1969je}
  J.~E.~Paton and H.~M.~Chan,
\emph{Generalized veneziano model with isospin,}
  Nucl.\ Phys.\  B {\bf 10} (1969) 516.



\bibitem{Kawai:1985xq}
  H.~Kawai, D.~C.~Lewellen and S.~H.~H.~Tye,
\emph{A Relation Between Tree Amplitudes Of Closed And Open Strings,}
  Nucl.\ Phys.\  B {\bf 269} (1986) 1.


\bibitem{Berends:1981rb}
  F.~A.~Berends, R.~Kleiss, P.~De Causmaecker, R.~Gastmans and T.~T.~Wu,
  \emph{Single Bremsstrahlung Processes In Gauge Theories,}
  Phys.\ Lett.\  B {\bf 103} (1981) 124.

\bibitem{Kleiss:1985yh}
  R.~Kleiss and W.~J.~Stirling,
\emph{Spinor Techniques For Calculating P Anti-P $\to$ W+- / Z0 + Jets,}
  Nucl.\ Phys.\  B {\bf 262} (1985) 235.

\bibitem{Xu:1986xb}
  Z.~Xu, D.~H.~Zhang and L.~Chang,
\emph{Helicity Amplitudes for Multiple Bremsstrahlung in Massless Nonabelian
   Gauge Theories,}
  Nucl.\ Phys.\  B {\bf 291} (1987) 392.

\bibitem{Gunion:1985vca}
  J.~F.~Gunion and Z.~Kunszt,
 \emph{Improved Analytic Techniques For Tree Graph Calculations And The G G Q
   Anti-Q Lepton Anti-Lepton Subprocess,}
  Phys.\ Lett.\  B {\bf 161} (1985) 333.



\bibitem{Dixon:1996wi}
  L.~J.~Dixon,
\emph{Calculating scattering amplitudes efficiently,}
  arXiv:hep-ph/9601359.

\bibitem{Parke:1986gb}
  S.~J.~Parke and T.~R.~Taylor,
 \emph{An Amplitude for $n$ Gluon Scattering,}
  Phys.\ Rev.\ Lett.\  {\bf 56} (1986) 2459.

\bibitem{Berends:1987me}
  F.~A.~Berends and W.~T.~Giele,
\emph{Recursive Calculations for Processes with n Gluons,}
  Nucl.\ Phys.\  B {\bf 306} (1988) 759.



\bibitem{Witten:2003nn}
  E.~Witten,
 \emph{Perturbative gauge theory as a string theory in twistor space,}
  Commun.\ Math.\ Phys.\  {\bf 252} (2004) 189
  [arXiv:hep-th/0312171].

\bibitem{Britto:2004ap}
  R.~Britto, F.~Cachazo and B.~Feng,
 \emph{New recursion relations for tree amplitudes of gluons,}
  Nucl.\ Phys.\  B {\bf 715} (2005) 499
  [arXiv:hep-th/0412308].

\bibitem{Britto:2005fq}
  R.~Britto, F.~Cachazo, B.~Feng and E.~Witten,
 \emph{Direct proof of tree-level recursion relation in Yang-Mills theory,}
  Phys.\ Rev.\ Lett.\  {\bf 94} (2005) 181602
  [arXiv:hep-th/0501052].



\bibitem{Benincasa:2007xk}
  P.~Benincasa and F.~Cachazo,
\emph{Consistency Conditions on the S-Matrix of Massless Particles,}
  arXiv:0705.4305 [hep-th].



\bibitem{Metsaev:2007rn}
  R.~R.~Metsaev,
  \emph{Cubic interaction vertices for fermionic and bosonic arbitrary spin
  fields,}
  arXiv:0712.3526 [hep-th].





\bibitem{Curtright:1980yk}
  T.~Curtright,
\emph{Generalized Gauge Fields,}
  Phys.\ Lett.\  B {\bf 165} (1985) 304.


\bibitem{Bengtsson:1983pd}
A.~K.~Bengtsson, I.~Bengtsson and L.~Brink,
\emph{Cubic Interaction Terms For Arbitrary Spin,}
Nucl.\ Phys.\ B {\bf 227} (1983) 31.

\bibitem{Bengtsson:1983pg}
A.~K.~Bengtsson, I.~Bengtsson and L.~Brink,
\emph{Cubic Interaction Terms For Arbitrarily Extended Supermultiplets,}
Nucl.\ Phys.\ B {\bf 227} (1983) 41.

\bibitem{Bengtsson:1986kh}
  A.~K.~H.~Bengtsson, I.~Bengtsson and N.~Linden,
\emph{Interacting Higher Spin Gauge Fields on the Light Front,}
  Class.\ Quant.\ Grav.\  {\bf 4} (1987) 1333.





\bibitem{Metsaev:2005ar}
  R.~R.~Metsaev,
\emph{Cubic interaction vertices of massive and massless higher spin fields,}
  Nucl.\ Phys.\  B {\bf 759} (2006) 147


\bibitem{Savvidy:2008zy}
  G.~Savvidy,
\emph{Interaction of non-Abelian tensor gauge fields,}
  arXiv:0804.2003 [hep-th].







\end{thebibliography}
\end{document}